\documentclass[twocolumn]{openjournal}
\usepackage{xcolor}
\usepackage{textgreek}
\usepackage[utf8]{inputenc}
\usepackage[english]{babel}

\usepackage{hyperref}
\hypersetup{
    unicode, 
    colorlinks=true,
    linkcolor=linkcolor,
    citecolor=linkcolor,
    filecolor=linkcolor,
    urlcolor=linkcolor,
}
\usepackage{color,colortbl}
\definecolor{linkcolor}{rgb}{0.0,0.3,0.5}
\usepackage{tensind}
\tensordelimiter{?}
\DeclareGraphicsExtensions{.bmp,.png,.jpg,.pdf}
\usepackage{verbatim}
\usepackage[normalem]{ulem}
\usepackage{orcidlink}
\usepackage{soul}
\usepackage{amsmath}
\usepackage{subfigure}

\graphicspath{ {./figs/} }

\addto\extrasenglish{%

\newcommand{\aref}[1]{\hyperref[#1]{Appendix~\autoref{#1}}}
}

\newcommand{\degree}{^\circ}

\begin{document}
\title{Analysis and Implications of the Spatio-Spectral Morphology of the Fermi Bubbles}

\author{Ami Tank\orcidlink{0009-0004-2104-4931}}
\email{aminimeshtank@gmail.com}
\affiliation{Department of Astronomy, Astrophysics and Space Engineering, Indian Institute of Technology Indore, Indore 453552, India }

\author{Roland Crocker\orcidlink{0000-0002-2036-2426}}
\email{rcrocker@fastmail.fm}
\affiliation{Research School of Astronomy \& Astrophysics, Australian National University, 233 Mt Stromlo Rd, Stromlo, ACT 2611, Australia}

\author{Mark R. Krumholz\orcidlink{0000-0003-3893-854X}}
\email{mark.krumholz@anu.edu.au}
\affiliation{Research School of Astronomy \& Astrophysics, Australian National University, 233 Mt Stromlo Rd, Stromlo, ACT 2611, Australia}

\begin{abstract}
The Fermi Bubbles are $\gamma$-ray structures extending from the center of the Milky Way to $\pm 50\degree$ Galactic latitude that were discovered in data obtained by the \textit{Fermi}/LAT instrument. Their origin and power source remain uncertain.
%
%An important feature of the Fermi Bubbles is that their $\gamma$-ray surface brightness is roughly constant  with respect to Galactic latitude. 
%
To help address this uncertainty, here we use a template-free reconstruction of ten years of all-sky \textit{Fermi}/LAT data provided by \cite{platz_multi-component_2023} to carry out a pixel-by-pixel spectral analysis of the Bubbles. We recover the position-dependent spectral shape and normalization that would be required for parent proton or electron cosmic ray populations to produce the Bubbles' observed $\gamma$-ray spectra.
%Here we use a template-free reconstruction of ten years of \textit{Fermi} LAT data provided by \cite{platz_multi-component_2023}.
%
%We perform a pixel-by-pixel fit of this data using \textbf{Naima} in order to recover the spatial variation of the normalisation and shape parameters describing proton or electron parent cosmic ray populations that could be responsible for the observed $\gamma$-ray spectral morphology via either the hadronic or leptonic channels, respectively.
%
%In this work, we perform a pixel-by-pixel fit of the gamma-ray emission via both these channels, calculated using a Python package called \textbf{Naima}, to a template-free reconstruction of ten years of \textit{Fermi} LAT data provided by \cite{platz_multi-component_2023}. For the ISRF, we adopt a model developed by \cite{popescu_radiation_2017}. We assume three different models for CRe and CRp, and infer their line-of-sight distribution required to reproduce the observed gamma-ray emission. 
We find that models in which the $\gamma$-ray emission is driven by either hadronic or leptonic processes can explain the data equally well. The cosmic ray population driving the emission must have either broken power-law or exponentially cut-off spectra, with break or cutoff energies that are almost constant with latitude but spectral indices below the break that
harden towards the Bubbles' southern tip. 
For the leptonic channel, reproducing the observed position-dependent $\gamma$-ray spectrum also requires a cosmic ray electron energy density that grows with distance from the Galactic plane and increases towards the edges of the Bubbles, implying either relativistic transport from the nucleus or \textit{in-situ} re-acceleration at the bubble edge. 
%
%This finding disfavors scenarios for the origin of the Bubbles where a population of cosmic ray electrons is accelerated near the Milky Way center and subsequently advected out to the extremities of the Bubbles.%
\end{abstract}

% Write your keywords here
\begin{keywords}
{Galactic center (565), Galactic cosmic rays (567), Gamma-ray sources (633), High-energy cosmic radiation (731)}
\end{keywords}

\maketitle

\section{Introduction}\label{sec:intro}

Observations of the Milky Way halo by the Large Area Telescope (LAT) onboard the \textit{Fermi} Gamma-ray Space telescope revealed two giant bubble-shaped structures extending $\sim$ 50$\degree$ above and below the Galactic plane. These structures were called the Fermi Bubbles (FBs) by their discoverers \citep{su_giant_2010}. The origin of the FBs is still an open question. Their morphology suggests that they originate from the Galactic Center (GC), and their sky location and morphology suggest they are plausibly the gamma-ray counterparts to the ``microwave haze'' first discovered \citep{finkbeiner_microwave_2004} in data from the Wilkinson Microwave Anisotropy Probe \citep[WMAP; also see][]{Dobler2012,Ade2013}.

Since the discovery of the FBs, structures coincident with or clearly morphologically related to them have also been detected at other wavelengths. The eROSITA telescope revealed two X-ray structures, now known as the eROSITA bubbles, extending up to 14 kpc above and below the Galactic plane \citep{predehl_detection_2020}; these encircle the FBs in projection and subsume previously-detected X-ray structures including the North Polar Spur (NPS) discovered by ROSAT \citep{snowden_rosat_1997} and the X- or hourglass-shaped structure detected near the GC \citep{bland-hawthorn_large-scale_2003} that is coincident with the edge of the FBs. 
In 2.3 GHz radio emission, \cite{carretti_giant_2013} detected two giant, linearly polarized lobes coincident with the FBs at low latitudes, but extending to greater heights ($\sim 60\degree$). There are also observations of both warm ionized \citep[e.g.,][]{Bordoloi17a}, cool atomic \citep[e.g.,][]{Noon23a}, and cold molecular \citep[e.g.,][]{Di-Teodoro20a, Cashman21a} material in or around the edges of the FBs, which may represent cool interstellar medium (ISM) that has been swept up by the same mechanisms responsible for producing the FBs.
\cite{sarkar_fermierosita_2024} provides a detailed review of the multi-wavelength observations near the FBs.

Despite the large volume of observations available, numerous basic questions about the FBs remain open. What microphysical mechanism generates the Bubbles' $\gamma$-ray emission? What activity provides the energy to inflate them in the first place? Relatedly, how old are the Bubbles? Where and how are the cosmic ray (CR) parents of the $\gamma$-rays accelerated? A successful model, along with answering these questions, should also be consistent with the full set of multi-wavelength observations. 
%(\cite{yang_unveiling_2018} provides a short review).

The question of mechanism arises because astrophysical production of $\gamma$-rays occurs via two primary channels: (a) the hadronic (p-p or proton) channel\footnote{In principle, this channel also encompasses heavier nuclei in either or both of the target and beam particles; in the absence of strong constraint on the abundance distribution for either of these, we ignore this complication. For Solar metallicity, the presence of heavy nuclei implies a $\mathcal{O}$(50\%) upwards correction \citep{Mori1997} to the efficiency of $\gamma$-ray production compared to a gas of pure hydrogen, which is small in comparison to other uncertainties, in particular, the target gas number density.} and (b) the leptonic (electron) channel. In the former, cosmic-ray protons (CRp) collide inelastically with ambient, thermal protons and produce neutral pions, which each further decay into two $\gamma$-ray photons. In the latter, $\gamma$-ray photons are produced via inverse Compton (IC) up-scattering\footnote{Leptonic $\gamma$-ray emission due to bremsstrahlung is also important in some astrophysical environments, but is not competitive with IC for the ISRF and gas number conditions  pertaining close to the FBs.} of the interstellar radiation field (ISRF) -- including the Cosmic Microwave Background (CMB) -- by cosmic-ray electrons (CRe). 
An appealing aspect of the leptonic channel is that synchrotron radiation by the same population of electrons could also generate the microwave haze emission \citep{su_giant_2010}. However, this is only suggestive, and we cannot at this point rule out the possibility that the $\gamma$-ray and microwave emission trace different underlying particle populations \citep[e.g.,][]{crocker_unified_2015}.

The question of mechanism is tightly tied to the question of the FBs' ultimate energy source. Models for this fall into two broad categories: those invoking AGN-like activity of the Milky Way's central super-massive black hole, and those that posit that the FBs are powered by supernovae produced by bursts of star formation near the GC. The link between these scenarios and the emission mechanism is via cooling times. CRe would cool in at most a few Myr due to IC and synchrotron emission (which are comparably important given the $\sim 10$ $\mu$G magnetic field found even at relatively high latitudes within the FBs -- \citealt{carretti_giant_2013}), so if FB emission is driven by CRe accelerated near the GC, those CRe must be transported outward at $\gg 1000$ km s$^{-1}$ to reach the FB edge before cooling. This is well above the typical speeds of supernova-driven winds \citep[e.g.,][]{Vijayan24a}, so in this scenario the CRe must either originate from a relativistic jet from the central black hole that was active in the last few Myr \citep[e.g.,][]{guo_fermi_2012}, or there must be some mechanism within the bubbles capable of re-accelerating a cooled electron population \citep[e.g.,][]{mertsch_fermi_2011,mertsch_fermi_2019}. By contrast, the cooling times for protons are much longer ($\sim$Gyr for ambient gas densities $\sim 10^{-2}$ cm$^{-3}$), easing the timescale constraint but conversely requiring a much larger CR energy density within the bubble to power the observed emission. This favors a scenario where the CRp population within the FBs is gradually built-up by the cumulative effects of supernova injection over $\gg$Myr timescales
\citep[e.g.,][]{crocker_fermi_2011, crocker_steady-state_2014, crocker_unified_2015}, though in principle CRp could also be supplied (and much more quickly) by AGN activity \citep[e.g.,][]{Mou2015}.

These cooling timescale arguments can be sharpened by considering variation in the emission across the FBs. The FBs
%Another important and constraining feature of the FBs is that they
have approximately uniform surface brightness and spectral shape \citep[]{su_giant_2010, ackermann_spectrum_2014}, which at first might seem difficult to reconcile with a leptonic model, since both the energy density and total intensity of the ISRF are expected to be non-uniform over such a large extent.
%
%In constrast,  
%the ISRF is spatially varying over the size scale of Bubbles, both in terms of total energy density and the overall spectral distribution as, in particular, the CMB comes to dominate the overall energy density at high latitudes (\autoref{fig:ISRF}).
%
%A natural question to ask is: what sort of spatial variation, if any, in the CRe distribution do we require to compensate the change in ISRF in order to produce a quasi spatially constant gamma-ray surface brightness and spectrum?
%This sort of question was addressed by
Despite this expectation, when
\citet{narayanan_latitude-dependent_2017} divided the FBs into five latitude bands and fit the spectra in each of them with a leptonic model, they found a surprisingly good fit 
%who performed a latitude-dependent analysis for the CRe distribution in a purely leptonic scenario considering five latitude bands. 
%
%In fact,  they found a surprisingly good fit to the FBs' gamma-ray spectral morphology 
even
with a constant  CRe spectrum.
However, they also found that such a model does not naturally reproduce the microwave haze, and that while it is possible to produce both the $\gamma$-ray and microwave emission with a single electron population, doing so requires significant fine-tuning of the position-dependent spectral shape.
%On the other hand, explaining the microwave haze emission via  synchrotron radiation, required a hardening of the CRe spectrum with latitude.
%
%They also found a solution where the CRe spectrum hardens in mid-latitudes while having a lower high energy cutoff\footnote{We do not explicitly address the microwave haze here and it does not constrain our modelling.}. 
Similarly, \cite{yang_spatially_2017} investigate a leptonic jet scenario by tracking the evolution of CRe using a 3D hydrodynamical simulation. They find that CRe in such a scenario develop a high energy cutoff due to fast synchrotron and IC cooling of the CRe when the jets were launched, but that the CRe population inside the FBs is then homogenized by the subsequent fast transport of CRe within the bubbles.
% While such a scenario can explain the relatively modest amount of spectral variation across the FBs, it does not naturally explain the uniform overall intensity, which requires that the energy density of CRe increase with latitude in order to compensate for the decrease in ISRF strength.% 
They also find that the CRe have a slight gradient towards higher energies at higher latitudes, which is essential to compensate for the decrease in the ISRF strength. 

However, all of the analyses to date have made use of relatively coarse representations of the variation of $\gamma$-ray emission across the FBs, for example simply taking means in latitude slices. As we discuss below, the combination of several additional years of \textit{Fermi}/LAT data with improved analysis techniques have yielded data sets with considerably better spatial and spectral resolution. This situation prompts us to revisit the question of what we can learn about the underlying particle population from a spatially-resolved analysis of the FBs' $\gamma$-ray spectra. 
%In this work, we present a data-driven and largely model-agnostic analysis of the spectral morphology of the Bubbles.
%
In this paper we perform a pixel-by-pixel analysis over $\approx 20,000 $ pixels covering the FBs, obtaining position-dependent constraints on the properties of the particle populations that must be present to explain the observed $\gamma$-ray spectra under either a leptonic or hadronic scenario\footnote{The dataset is released at \url{https://doi.org/10.5281/zenodo.19675422}}.

%of the gamma-ray data in order to recover position-dependent parent CR distribution parameters for both leptonic and hadronic microphysical scenarios.
%
%We try three generic parametric models for the CR distribution -- pure power law, broken power law, and exponentially cut-off power law -- and fit to the data using the \textbf{Naima} Python package \citep{naima}.
%
%Our analysis covers a total of 37,714 pixels (each with a solid angle of \SI{6.5e-05}{sr}) covering a total solid angle of $\sim\SI{2.5}{sr}$, allowing us to infer the distribution of CRp and CRe along the line of sight that would reproduce the observed gamma-ray emission in each pixel. For the gamma-ray data, we use data obtained in a reconstruction developed by \cite{platz_multi-component_2023}.
%
%For our leptonic scenario analysis, we adopt the ISRF model developed by \cite{popescu_radiation_2017}.

The remainder of this paper is organized as follows. In \autoref{sec:Data} we discuss the $\gamma$-ray data reconstruction and the ISRF model data that we use. In \autoref{sec:methodology} we outline our methodology and discuss the assumptions we make about the structure of the FBs for this work. In \autoref{sec:results} and \autoref{sec:Discussion} we set out our results and discussion, respectively. 

\section{Data} \label{sec:Data}

%To model the CRp and CRe population, one needs the following data: ambient gas density $n_H$ (for the proton channel), ISRF distribution (for the electron channel), and the gamma-ray spectrum of the FBs. 
%
%Given that $n_H$(and, in particular, its spatial variation) is not well measured in the Bubbles, for hadronic scenarios, we simply normalise our results to a fiducial $n_H = 0.01$ cm$^{-3}$. 

Our analysis relies on measurements and models for the all-sky $\gamma$-ray spectrum and the interstellar radiation field; we discuss these in \autoref{ssec:gamma_ray} and \autoref{ssec:isrf}, respectively.

\subsection{$\gamma$-ray data}
\label{ssec:gamma_ray}

\begin{figure*}
\centering
\includegraphics[width=0.9\linewidth]{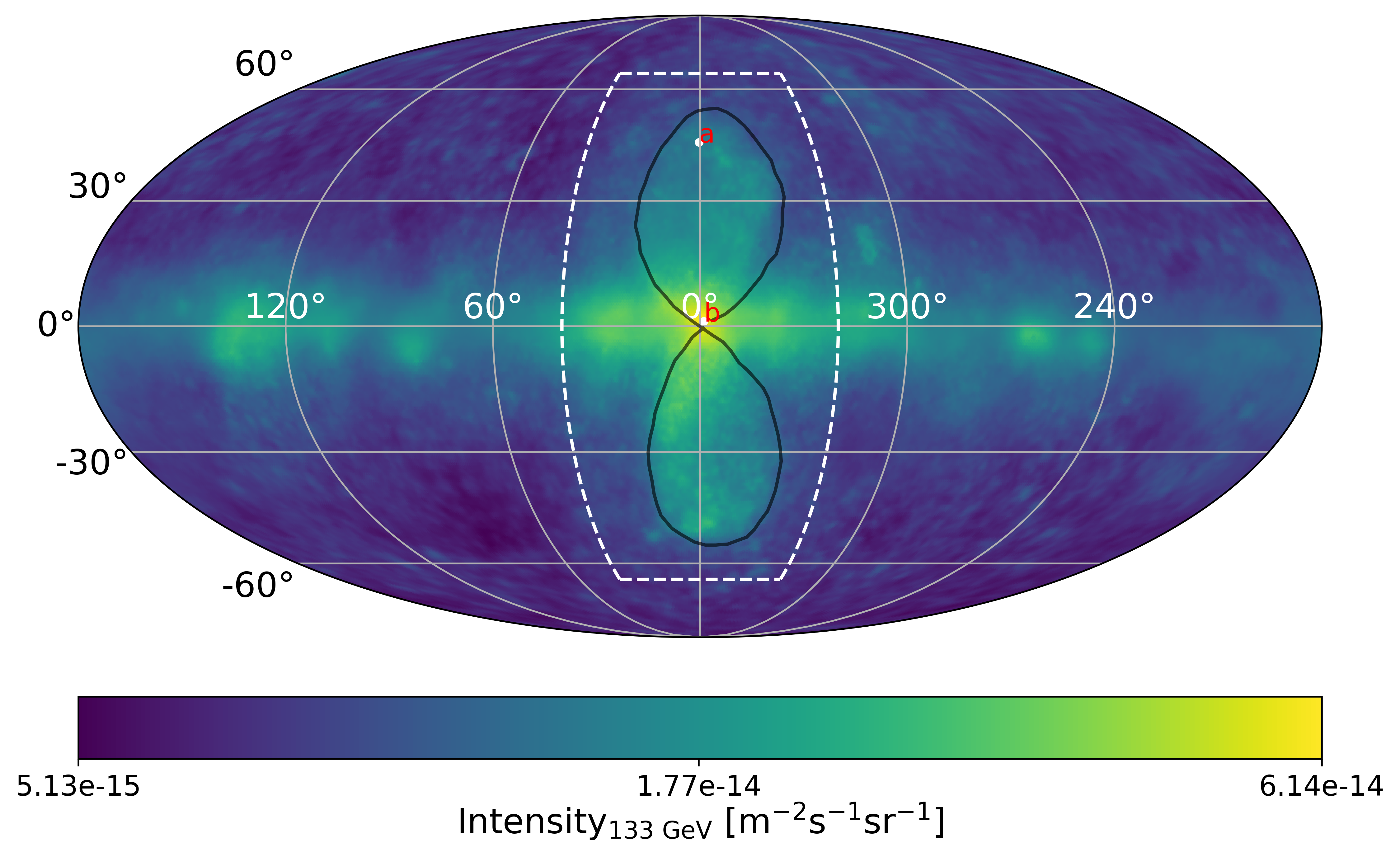}
\caption{Template-free reconstruction of the non-dust associated all-sky diffuse $\gamma$-ray photon flux, $I^\mathrm{nd}$, taken from the M2 model of \citet{platz_multi-component_2023}; the data shown are for the energy bin centered at 133 GeV. The data are shown in Mollweide projection using Galactic coordinates. The colorbar is plotted on a log scale. The black line is a by-eye tracing of the FBs' apparent edges, and the white dashed line shows the region $-40^\circ < \ell < 40^\circ$, $-65^\circ < b < 65^\circ$ on which we focus in this work. The two white points labeled `a' and `b' are chosen to show example results at different points in the sky.\\}
\label{fig:gamma_ray_data}
\end{figure*}

We based our analysis on the all-sky $\gamma$-ray maps derived by \citet{platz_multi-component_2023}. These authors apply a Bayesian variational inference framework to 10 years of \textit{Fermi}/LAT data, which they use to obtain
%and based on phenomenological modeling of the emission components of the gamma-ray sky. These authors take the time-averaged, total gamma-ray photon flux density, as a function of photon origin direction and energy, as their signal of interest (in units of (s m$^2$ GeV sr)$^{-1}$). 
%
%With respect to energy, the data are sorted into 11 logarithmically spaced energy bins ranging from 0.56 to 316 GeV (4 bins per decade of energy), of which we discard the lowest energy bin\footnote{\citet{platz_multi-component_2023} urge caution in using the two lowest energy bins due to potential instrument response function mismodeling. 
%
%In our analysis we have therefore decided to discard the first energy bin.}. 
%
%With respect to direction, the sky is pixelised based on the HEALPix `RING' scheme with an \textit{nside} value of 128. This divides the entire sky into 196608 pixels, each with a solid angle of \SI{6.5e-05}{sr}. 
%
%A pixel-wise reconstruction of the integral signal is constructed by summing, in a fixed direction, over the energy bins and dividing by the solid angle of each sky pixel (in units of (m$^2$ s sr)$^{-1}$).  
%\citet{platz_multi-component_2023} present
two different decompositions of the all-sky diffuse photon flux density as a function of direction $I$: a template-free model M1, and a template-informed model M2.
%
%Both of these models allow for point sources (PS) in addition to a diffuse emission component. 
%
%A separate component for the isotropic diffuse background emission is not included in the models, and is hence absorbed in this diffuse emission component. 
%
%These models are built within a hierarchical Bayesian framework where each component is modeled as a stochastic field with spatial and spectral correlations. The model incorporates physically motivated priors such as smoothness, isotropy and separability across spatial and spectral dimensions, but does not assume fixed spectral indices or spatial morphologies. Instead, the generative models are guided by weak priors from the literature -- such as typical gamma-ray spectral indices and correlations between point source flux and spectral hardness -- thus retaining the flexibility to derive the spectra directly from the data.
%While the M1 model does not incorporate any spatial template, the M2 model 
The latter divides diffuse emission into two components:
%introduces a dust-correlated template and further subdivides the diffuse component into two. The first component, 
$I^\mathrm{dust}$, 
%is template-informed and 
which
uses the \textit{Planck} 545 GHz thermal dust emission map as a template,
%which correlates strongly with the denser gas that provides target nuclei for gamma-ray production by the Galactic plane CR hadrons. The second component, 
and
$I^\mathrm{nd}$, which is template-free and therefore accounts for the processes which are uncorrelated with the dust map.
We expect the dust-associated component to trace processes such as $pp$ emission from the Galactic disk, which strongly correlates with the location of dense interstellar gas and thus dust, while the nd (no dust) component should capture processes
such as IC scattering from CRe or hadronic emission from regions not traced by the dust map. 
 
Most importantly for our purposes, while the FBs are visible in both M1 and M2 reconstructions, in M1 they are somewhat obscured by the strong foreground (and background) emission from the Galactic disk and halo. 
In contrast, the $I^\mathrm{dust}$ component in M2 effectively separates out the foreground largely attributable to the emission from the quasi-uniform `sea' of Galactic plane CR hadrons colliding with denser gas, revealing the FBs clearly in the $I^\mathrm{nd}$ component. 
For our analysis we therefore employ the template-free component, $I^\mathrm{nd}$, of the M2 model\footnote{The model is available at \url{https://doi.org/10.5281/zenodo.7970865}}. The map we use pixelizes the sky using the \texttt{HEALPix} \texttt{RING} pixelization scheme \citep{Gorski05a} with an \texttt{nside} parameter of 128; it therefore decomposes the sky into 196608 pixels of $6.5\times 10^{-5}$ sr each, and within each pixel the map provides the photon flux in 11 logarithmically-spaced energy bins from $0.56 - 316$ GeV; it also provides corresponding uncertainties on these fluxes. Following the advice of \citet{platz_multi-component_2023}, we discard the lowest energy bin from our analysis due to large systematic uncertainties arising from poor knowledge of the low-energy instrumental response function.

%Fig.~\ref{fig:gamma_ray_data} 
\autoref{fig:gamma_ray_data} shows the $\gamma$-ray sky traced by the $I^\mathrm{nd}$ component of M2 for an energy bin centered (geometrically) at 133 GeV. The black contour in this figure shows our by-eye tracing of the apparent boundary of the FBs which we use in some of our analyses.
%and when discussing some of our  results. 
%
%We calculated the luminosity of FBs from these data and obtained a value of $\sim\SI{4e+37}{erg s^{-1}}$. %

\subsection{The interstellar radiation field}
\label{ssec:isrf}

\begin{figure}
\centering
\includegraphics[width=0.9\linewidth]{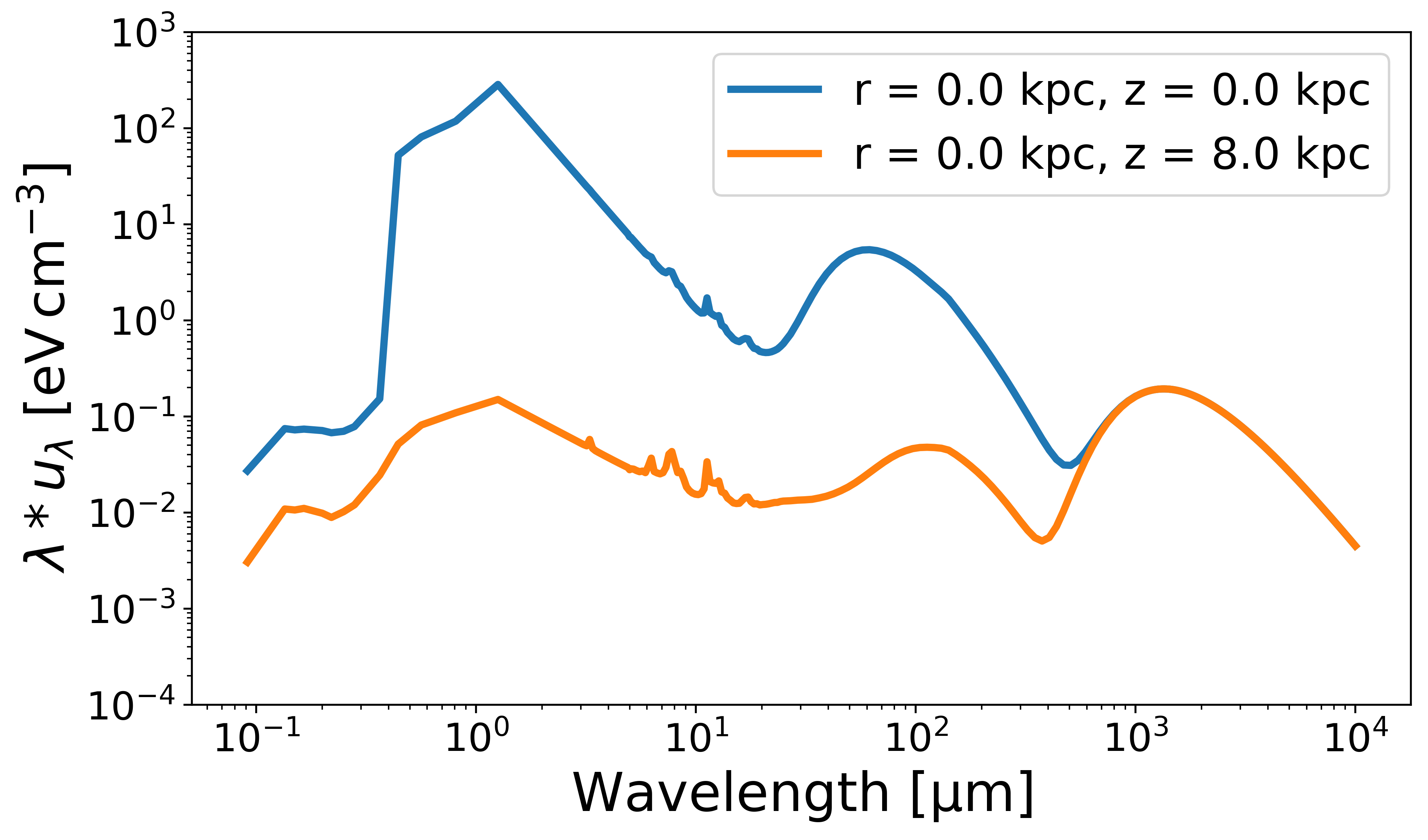}
\caption{Our model of the specific energy density $u_\lambda$ of the ISRF (including stellar, dust, and CMB components) as a function of wavelength $\lambda$, drawn primarily from \citet{popescu_radiation_2017}, at two sample positions: the Galactic center ($r=0$, $z=0$; blue) and at a point 8 kpc above the Galactic center ($r=0$, $z=8$ kpc; orange).\\}
\label{fig:ISRF}
\end{figure}

In leptonic models for the FBs, emission depends on the ISRF, and we therefore also require a model for the ISRF as a function of position. For this purpose we use the model produced by \citet{popescu_radiation_2017}, which is an axisymmetric radiative transfer (RT) model derived using the \textit{COBE}, \textit{IRAS} and \textit{Planck} maps of the all-sky emission.\footnote{The model data are available at \url{https://doi.org/10.26093/cds/vizier.74702539}.} The model is spatially discretized onto a cylindrical grid of $23\times 22$ points covering radii $r \in (0,24)$ kpc and heights $z\in (0,10)$ kpc, with symmetry assumed about the $z=0$ plane. At each point it describes the specific energy density of the radiation field $u_\lambda$ as a function of wavelength $\lambda$ as the sum of 
%the cosmic microwave background (a spatially-uniform blackbody field with temperature $T=2.73$ K),%
a direct starlight component covering wavelengths from 91 to 5000 nm (discretized into 15 bins), and a dust-reprocessed component from 3 $\mu$m to 1 mm (discretized into 120 bins); to avoid an artificial gap between the components, we extrapolate the stellar component to longer wavelengths by fitting a powerlaw to the three longest wavelength points, following \citet{niederwanger_consequence_2019}. The total ISRF also contains a contribution from the cosmic microwave background (CMB), a spatially uniform blackbody field with temperature $T=2.73$ K. We show the total ISRF 
%we produce via this procedure% 
at two sample points in \autoref{fig:ISRF}.
%ISRF is divided into direct starlight and dust re-radiated starlight components. The stellar component ranges from 91 nm (UV) to 5000 nm (NIR) discretised into 15 wavelength bins. The dust component ranges from 3 $\mu$m to 1 mm discretised into 120 wavelength bins. 

%The spatial ISRF model data are provided on a cylindrical grid with 23 points in the radial direction (\textit{r}) and 22 points in the vertical direction (\textit{z}) with variable spacing\footnote{The ISRF data is available here: \url{https://doi.org/10.26093/cds/vizier.74702539}}. The grid extends to 24 kpc radially and to 10 kpc vertically. The stellar component dominates the dust component up to $\sim$ 10$\mu$m. We extrapolate the stellar component to longer wavelengths by fitting a simple power law at the last three available data points, following the approach of \cite{niederwanger_consequence_2019}. 

%In addition to the stellar and dust emission components defined by \citet{popescu_radiation_2017}, we also account for the contribution from the CMB  to the total ISRF. The energy density of CMB is calculated using the blackbody spectrum at 2.73 K, although we directly use the inbuilt function in \textbf{Naima} to account for the CMB contribution. 
%
%\autoref{fig:ISRF} shows --   at two spatial locations -- the combined wavelength-multiplied spectral energy density per wavelength of the ISRF calculated after extrapolating the stellar component to longer wavelengths, and adding it to the dust component and to the CMB component  (the Galactic centre and 8 kpc above the Galactic center).

As we describe below, our analysis requires that we be able to compute the ISRF at arbitrary spatial positions. For positions within the range of $(r,z)$ covered by the \citeauthor{popescu_radiation_2017}~grid, we use bilinear interpolation on the grid they provide. We also require the ISRF at $z > 10$ kpc, beyond the provided grid, for which purpose we assume that the starlight and dust components of the radiation energy density follow a powerlaw functional form in $z$ at each radius $r$. For $r < 4$ kpc, where the \citeauthor{popescu_radiation_2017}~model grid behaves smoothly at large $z$, we perform a powerlaw fit to the three data points with the largest $z$ to generate $u_\lambda(r,z)$ for $z > 10$ kpc. For $r > 4$ kpc the behavior of the radiation field at large $z$ is more complex and a powerlaw fit is poorly-behaved, so we instead simply assume a $z^{-1}$ scaling, typical of the best-fit values at smaller $r$, and extrapolate from the largest $z$ value at each $r$ under this assumption. Note that, while formally necessary for us to cover our full target region up to Galactic latitudes $b=\pm 65^\circ$, these extrapolations have limited impact because almost all of the volume of the FBs lies below $z=10$ kpc.

%Note that for our analysis we require ISRF values beyond 10 kpc in the vertical direction, beyond the data range tabulated by \citet{popescu_radiation_2017}. We therefore extrapolate the spectral energy density of the combined (stellar + dust) ISRF to larger \textit{z} values. The energy density versus z plot is a smooth curve for $r<4$ kpc but for $r>4$ kpc there are bumps at the last three data points which are to be used for extrapolation (\autoref{extrapolate_ISRF} in \autoref{appendix}). Hence, we extrapolate the energy density to larger `z' values by fitting a power law to the last three data points for $r<4$ kpc and by assuming a z$^{-1}$ dependence of the energy density for $r>4$ kpc. %by fitting a power law to the last three available data points for smaller \textit{r} values, and by assuming a z$^{-1}$ dependence of the energy density for larger \textit{r} values. 
%by fitting a z$^{-1}$ power law to the last three spatial sampling points. 
%
%In addition, given the data is available only at a limited number of sampling points,  we use linear interpolation to calculate the ISRF at positions interior to the spatial points sampled by \citet{popescu_radiation_2017}. 

Finally, it is worth noting that many previous studies of the FBs \citep[e.g.,][]{yang_fermi_2013, yang_fermi_2022,narayanan_latitude-dependent_2017, yang_spatially_2017, herold_hard_2019}, relied on estimates of the spatial variation of the ISRF from \texttt{GALPROP}\footnote{\url{https://galprop.stanford.edu}}.  \citet{popescu_radiation_2017} point out that their ISRF model differs systematically from the \texttt{GALPROP} model v.54 (the pre-2017 model), particularly in the UV and MIR range. 
The \texttt{GALPROP} model of this generation obtains the stellar and dust ISRF components from star counts and gas measurements but \citet{popescu_radiation_2017} point out that this leads to an inconsistency with direct, inner Galaxy measurements by  \textit{COBE/IRAS/Planck};
%They point out that the predictions of GALPROP model are not consistent with the data observed by \textit{COBE/IRAS/Planck} as the GALPROP model assumes the stellar and dust distribution to be known from stellar counts and gas measurements wheres 
\cite{popescu_radiation_2017} optimize their stellar and dust ISRF component models such that they reproduce the \textit{COBE}, \textit{IRAS} and \textit{Planck} maps. Consequently, their estimate of the radiation field should be more accurate near the Galactic center, an important region for our application.
%
%They also point out that the difference in the MIR emission is due to the difference in the treatment of the local heating of dust by radiation from young stars in parent molecular clouds of star-forming regions, and also due to the difference in the treatment of the diffuse component of MIR emission. 

%In addition to the stellar and dust radiation fields, there is also a contribution from the cosmic microwave background (CMB). This is taken care of by an inbuilt function in the \textbf{Naima} package which we use to calculate the gamma-ray emission. This is further discussed in section~\ref{sec:methodology}.%

\section{Methodology} \label{sec:methodology}

Our goal in this section is to fit every pixel in the FBs with a range of potential models for the high-energy particle population driving the $\gamma$-ray emission. Given our aim, we restrict our attention to the region between $-65\degree < b < 65\degree$ and $-40\degree < \ell <40\degree$, where $(\ell,b)$ are the usual Galactic longitude and latitude coordinates, indicated by dashed white line in \autoref{fig:gamma_ray_data}. We first describe how we compute the auxiliary quantities necessary for fits in \autoref{ssec:isrf_density} and then detail the fitting method itself in \autoref{ssec:models}.

\subsection{ISRF and gas density estimation}
\label{ssec:isrf_density}

%We convert the units of the gamma-ray data from m$^{-2}$sr$^{-1}$s$^{-1}$ to GeVs$^{-1}$cm$^{-2}$ by multiplying each data point with the central value of the energy bin and with the solid angle of each pixel (\SI{6.5e-05}{sr}) rendering it compatible with the units used in \textbf{Naima} (see below). 
%
Leptonic models for the FBs require knowledge of the ISRF, which the models of \citet{popescu_radiation_2017} provide to us as a function of cylindrical $(r,z)$ coordinates. The observations, by contrast, provide the integrated intensity along a line of sight characterized by a pixel location in Galactic coordinates $(\ell,b)$, which passes through a range of values of $(r,z)$. It is therefore necessary to use the \citeauthor{popescu_radiation_2017} models to estimate the mean radiation field along each line of sight.

In order to do so, we assume that the FBs are figures of rotation generated by rotating the by-eye tracing of their apparent boundary shown in \autoref{fig:gamma_ray_data} about the $z$-axis. We construct two such curves $R(z)$, one for the Northern Bubble and one for the Southern Bubble, describing the cylindrical radius $r$ at each height $z$ that marks the edge of the FBs (and which we use to mark the apparent boundary of the FBs in later plots). Each observed coordinate $(\ell,b)$ then defines a unique ray, the intersection of which with the FB bounding surface will define the range of ISRFs that contribute to the emission seen along that ray. To express the problem algebraically, we adopt a Cartesian coordinate system centered on the Galactic Center, and with the Sun located at position $\mathbf{r}_\odot = (d_\mathrm{GC},0,0)$, where $d_\mathrm{GC} = 8.2$ kpc \citep{Bland-Hawthorn16b} is the distance to the Galactic Center. In this coordinate system, the ray corresponding to a particular line of sight $(\ell,b)$ in Galactic coordinates takes the form $\mathbf{r}_{(\ell,b)} = \mathbf{r}_\odot + s \hat{\mathbf{n}}_{(\ell,b)}$, where
\begin{equation}
    \hat{\mathbf{n}}_{(\ell,b)} = \left(-\cos \ell \cos b, \sin\ell\cos b, \sin b\right),
    \label{eq:ray_eqn}
\end{equation}
is a unit vector pointing in the direction of the line of sight corresponding to Galactic coordinates $(\ell,b)$ and $s$ is the distance from the Sun along the line of sight. Then the distances along a given line of sight at which it enters and exits the FBs are given implicitly by the solutions to
\begin{equation}
    (d_\mathrm{GC} + s \hat{n}_{x,(\ell,b)})^2 + s^2 \hat{n}_{y,(\ell,b)}^2 = R^2(s \hat{n}_{z,(\ell,b)}),
\end{equation}
where $R(z)$ is our function tracing out the FB edge. For lines of sight that pass through the FBs this equation has two real solutions $s_\pm$, which we can then use to define the Cartesian positions at which the ray enters and exits the FBs, which we denote $\mathbf{r}_{\pm,(\ell,b)}$. We then define the mean ISRF experienced along each line of sight by taking the average along the ray, i.e., we define
\begin{equation}\label{mean_isrf}
    \langle u_\lambda\rangle_{(\ell,b)} = \frac{1}{d_{(\ell,b)}} \int_0^{d_{(\ell,b)}} u_\lambda(\mathbf{r}_{-,(\ell,b)} + s'\hat{\mathbf{n}}_{(\ell,b)}) \, ds', 
\end{equation}
where $d_{(\ell,b)} = |\mathbf{r}_{+,(\ell,b)} - \mathbf{r}_{-,(\ell,b)}|$ is the distance between the entry and exit points, and we evaluate the function $u_\lambda(\mathbf{r})$ giving the radiation specific energy density at position $\mathbf{r}$ by converting the position to cylindrical coordinates and interpolating on the table of $u_\lambda$ as a function of $(r,z)$ supplied by \citeauthor{popescu_radiation_2017}

%We consider a vector along the line of sight from the Sun to each pixel on the plane and determine the points at which it crosses the surface of the Bubbles as traced by R(z) and use them to calculate the depth of the FBs along the line of sight which is used later to compute the energy density of the CRe. We then assign a mean ISRF energy density $\langle u_\lambda\rangle$ to each pixel by averaging the ISRF computed at multiple points along the line of sight within the Bubble according to \autoref{mean_isrf}.} 

%The stellar component of the ISRF has bumps at some points along the radial direction. This is an artefact of the RT calculations in the ISRF model  \citep[Fig.6 in][]{popescu_radiation_2017}. Since we do not want this artefact to influence our final results, we average the combined (stellar + dust) ISRF along the line of sight over a radial distance of 2 kpc from the assumed plane on both sides. We then assign these averaged ISRF values to each pixel and use them as the energy densities of the seed photon fields in \textbf{Naima}. 

Hadronic models of the FBs in turn require knowledge of the gas density as a function of position within the FBs. Unfortunately this is poorly known and highly model-dependent. We cannot, for example, simply adopt models for the density structure of a steady-state circumgalactic medium \citep[e.g.,][]{Stern19a}, since in hadronic models the FBs themselves are assumed to be inflated by the combined pressure of non-thermal particles and hot gas, and thus are not expected to follow the density of the surrounding CGM. For this reason, we will adopt a fiducial uniform number density of H nuclei $n_\mathrm{H} = 10^{-3}$ cm$^{-3}$, roughly the value predicted near the tops of the bubbles by some recent models \citep[e.g.,][]{Tourmente2025}. Because this is so uncertain, however, we will report all quantities for hadronic models normalized to this density.

\subsection{Model fits}
\label{ssec:models}
%We use the \textbf{Naima} Python package to forward model the non-thermal radiation from relativistic particle distributions with prescribed parameters (\cite{naima}) and to calculate the spectral energy distribution (SED) of gamma-ray emission via the proton and electron channels. 
We use the \texttt{Naima} Python package \citep{naima} to forward model the non-thermal $\gamma$-ray emission from relativistic proton and electron distributions with prescribed spectral parameters. For hadronic models we use the \texttt{Naima} \texttt{PionDecay} class, which calculates the $\gamma$-ray spectrum resulting from the interaction of a relativistic CRp population with ambient protons using the parameterization of \cite{kafexhiu_parametrization_2014}; as noted above, for hadronic models we normalize to a fiducial proton density $n_\mathrm{H}=10^{-3}$ cm$^{-3}$. 
For leptonic models, we use the \texttt{Naima} \texttt{InverseCompton} class to calculate the $\gamma$-ray spectrum produced by IC scattering of seed photon fields by an assumed CRe distribution. This calculation uses the differential cross-section provided by \cite{khangulyan_simple_2014} for blackbody photon fields and a cross-section developed by \cite{aharonian_compton_1981} for non-thermal photon fields. We use the former capability to compute $\gamma$-ray production by IC scattering of the CMB, and the latter to compute production by IC scattering of the ISRF, using the mean ISRF energy density $\langle u_\lambda\rangle$ for each pixel derived as described in \autoref{ssec:isrf_density}. 
%The seed photon fields are to be given as a list in the following format: [name, energy, energy density]. We give 135 such lists (one for each wavelength in the \cite{popescu_radiation_2017} ISRF data) as seed photon fields and use the inbuilt function for the CMB field. 

%We give a list of previously discussed ISRF as seed photon fields. We also use an inbuilt function in \textbf{Naima} for the CMB photon field, which assumes a temperature of 2.73 K and an energy density of 0.26 eV cm$^{-3}$.%

We consider three models for the energy distributions of parent CRp and CRe populations: (i) power-law (PL), (ii) power-law with an exponential cut-off (EPL), and (iii) broken power-law (BPL). The functional forms and free parameters for these models are 
\begin{enumerate}
    \item[PL:] $N(E) = A(E/E_{0})^{-\alpha}$, where $\alpha$ is a free spectral index.
    \item[EPL:] $N(E) = A(E/E_{0})^{-\alpha}\exp(-E/E_\mathrm{cut})$, where $\alpha$ is a free spectral index and $E_\mathrm{cut}$ is a free cut-off energy.
    \item[BPL:] $N(E) = A(E/E_{0})^{-\alpha_{1}}$ for $E<E_{\rm break}$ and $N(E) = A(E_{\rm break}/E_{0})^{\alpha_{2}-\alpha_{1}}(E/E_{0})^{-\alpha_{2}}$ for $E>E_{\rm break}$, where $\alpha_1$ and $\alpha_2$ are free spectral indices and $E_{\rm break}$ is a free break energy.
 \end{enumerate}
In these expressions, $N(E)$ is the differential number density of CR particles as a function of energy $E$, $E_0 = 10$ GeV is an arbitrary reference energy, and $A$ is a free amplitude. Thus the PL model has two free parameters ($A$ and $\alpha$), the EPL model has three ($A$, $\alpha$, $E_\mathrm{cut}$), and the BPL model has four ($A$, $\alpha_1$, $\alpha_2$, $E_\mathrm{break}$).

We find the best fit model parameters for each pixel by fitting the $\gamma$-ray spectrum calculated using \texttt{Naima} to the data using the \texttt{SciPy} \texttt{curve\_fit} routine, which uses non-linear least squares fitting.
%We find the best fit model parameters by fitting the computed SED to the gamma-ray data at each pixel.% 
We evaluate the goodness of fit of each model by computing the reduced chi-square value, given by $\chi^{2}_\mathrm{red}$ = $\chi^{2}$/$\nu$, where $\chi^2$ = $\sum_{j=1}^{10} (\Phi_{\mathrm{obs},j} - \Phi_{\mathrm{mod},j})^2/\sigma_j^2$, $\Phi_{\mathrm{mod},j}$ and $\Phi_{\mathrm{obs},j}$ are the model-predicted and observed (taken from the M2 non-dust component -- cf.~\autoref{ssec:gamma_ray}) photon fluxes in the $j$th energy bin (out of 10 total) for each pixel, $\sigma_j$ is the corresponding uncertainty, and $\nu$ is the number of degrees of freedom in the fit, given by the difference between the number of observations and the number of fitted parameters. However, this quantity should be interpreted with considerable caution, because the errors $\sigma_j$ are \textit{not} predominantly random measurement errors; instead they are dominated by uncertainties in the decomposition, which means that uncertainties in adjacent energy bins are highly-correlated rather than independent. Consequently, we cannot give the reduced $\chi^2$ its usual interpretation in terms of an absolute goodness of fit; we will instead simply use it to characterize relative goodness of fit between alternative models.

\begin{figure}
\centering
\includegraphics[width = 0.95\linewidth]{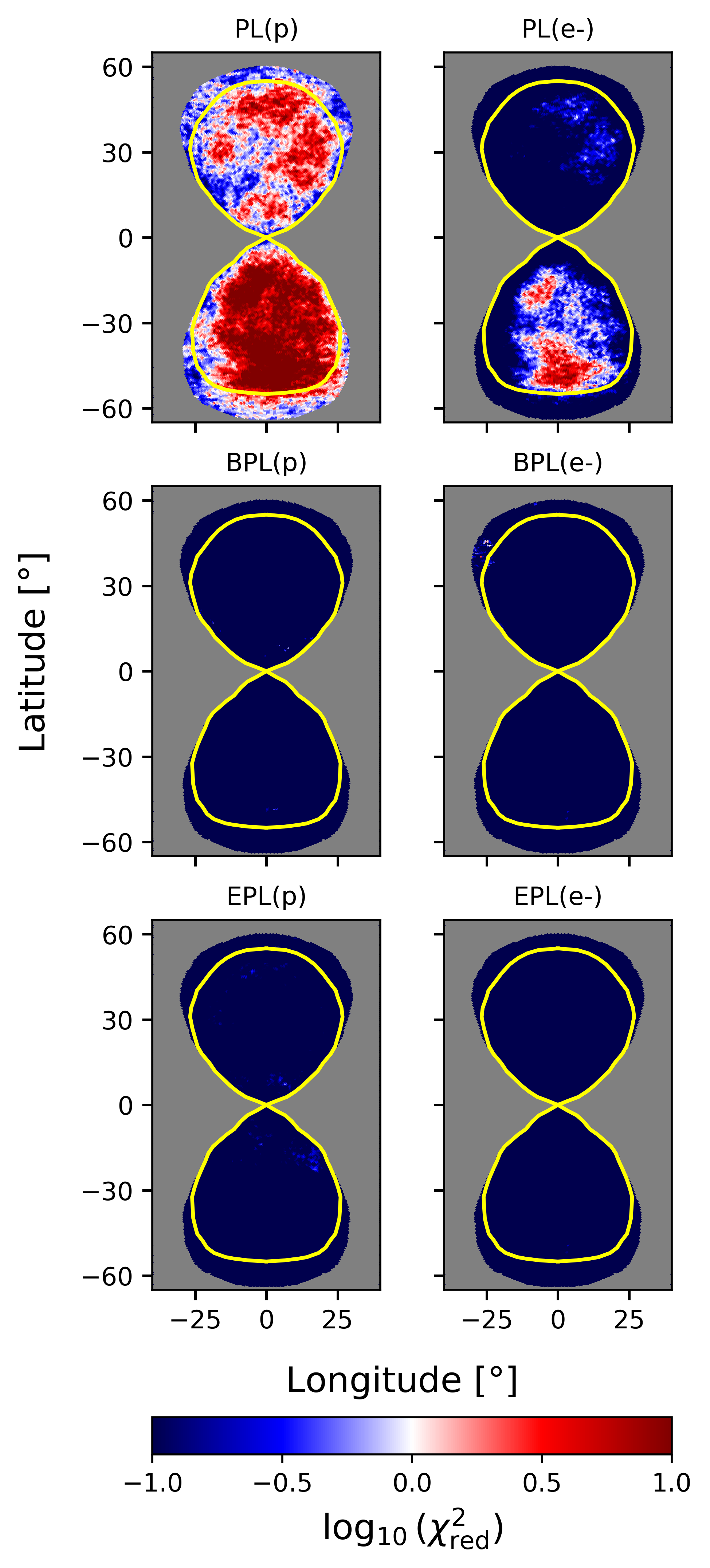}
\caption{The log of reduced $\chi^2$ for each model and source particle population (left panel for protons and right panel for electrons, PL, BPL, and EPL levels from top to bottom) for each pixel. The boundary of the FBs is marked by a yellow line, which is the same in each panel. We see that a power law model with a break or an exponential cutoff provides a better fit than a simple PL model in the FBs region.\\}
\label{fig:model_comparison}
\end{figure}

\begin{figure*}
\centering
\includegraphics[width=0.95\linewidth]{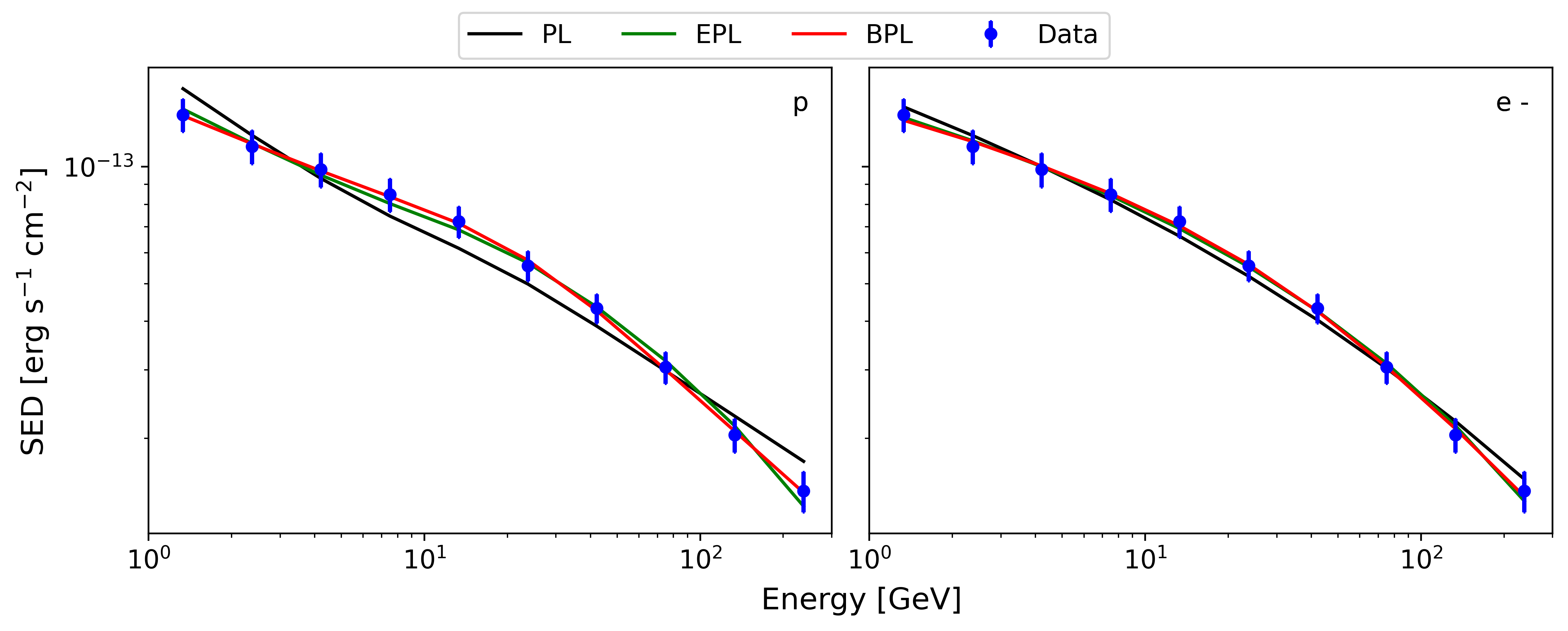}
\caption{Best-fit model spectra versus energy-binned data for the six different models evaluated at the point $b=45\degree$ and $l=0.38\degree$ (labeled `a' in \autoref{fig:gamma_ray_data}) within the FBs. The blue points with error bars show the $\gamma$-ray data, and the computed SED is shown by solid black (PL), green (EPL), and red (BPL) lines. The EPL and BPL models perform better than the PL models for both hadronic (left) and leptonic (right) cases.\\}
\label{fig:fit_comp}
\end{figure*}

\section{Results} \label{sec:results}

Having carried out fits to each pixel using each parent population model as described in \autoref{sec:methodology}, we are now in a position to examine what the fits tell us.

\subsection{Which models can explain the data?}

\begin{figure}
\centering
\includegraphics[width=0.95\linewidth]{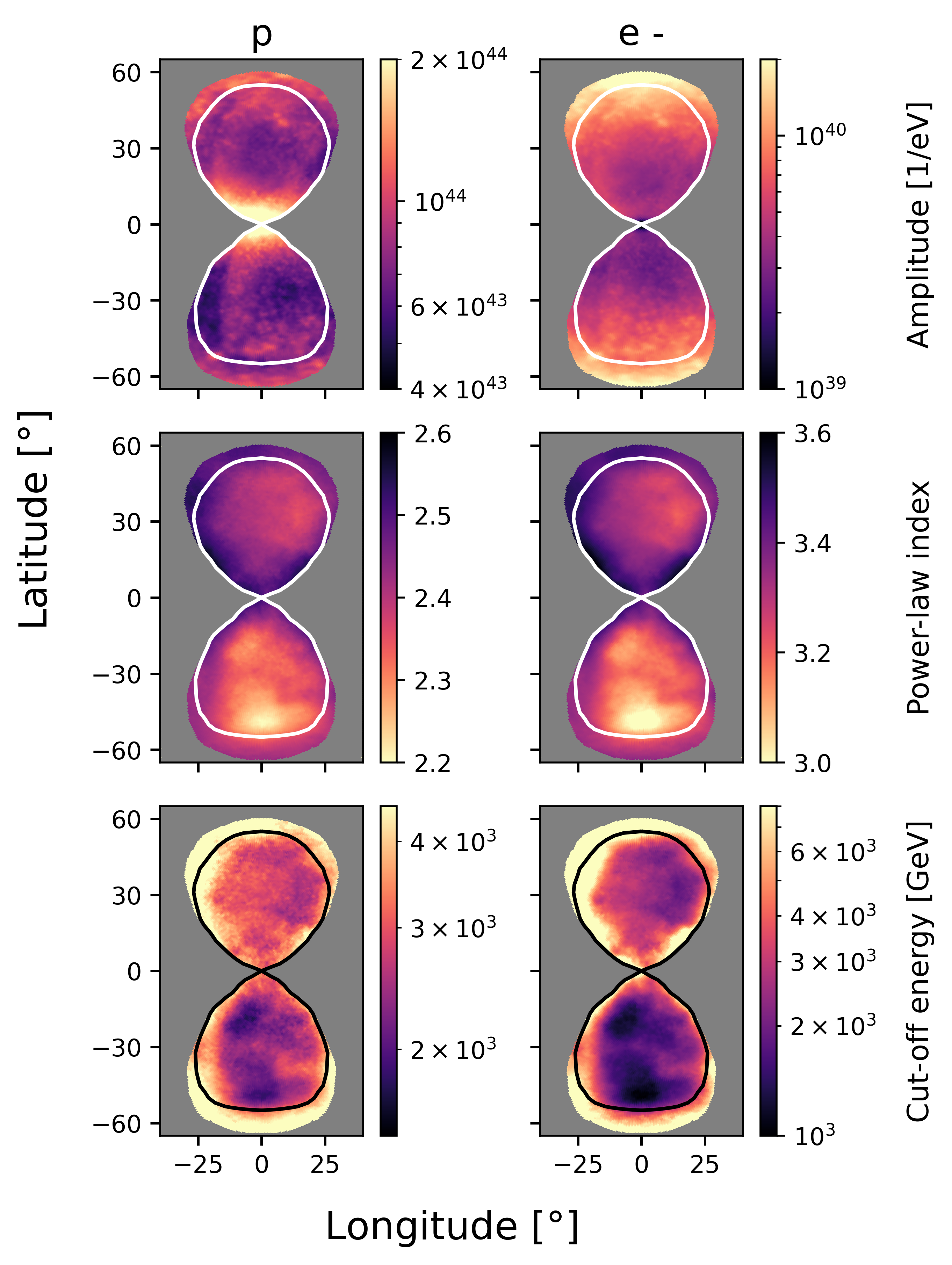}
\caption{Best-fit parameters for the EPL models (protons on the left and electrons on the right). The plots in the top panel shows the model amplitude and the middle panel shows the power-law index. We see a spectral hardening towards the Southern tip for both. The bottom panel shows the cut-off energy. It is of the order of a few TeV and does not vary much with latitude.\\}
\label{fig:EPL_model}
\end{figure}

\begin{figure}
\centering
\includegraphics[width=0.95\linewidth]{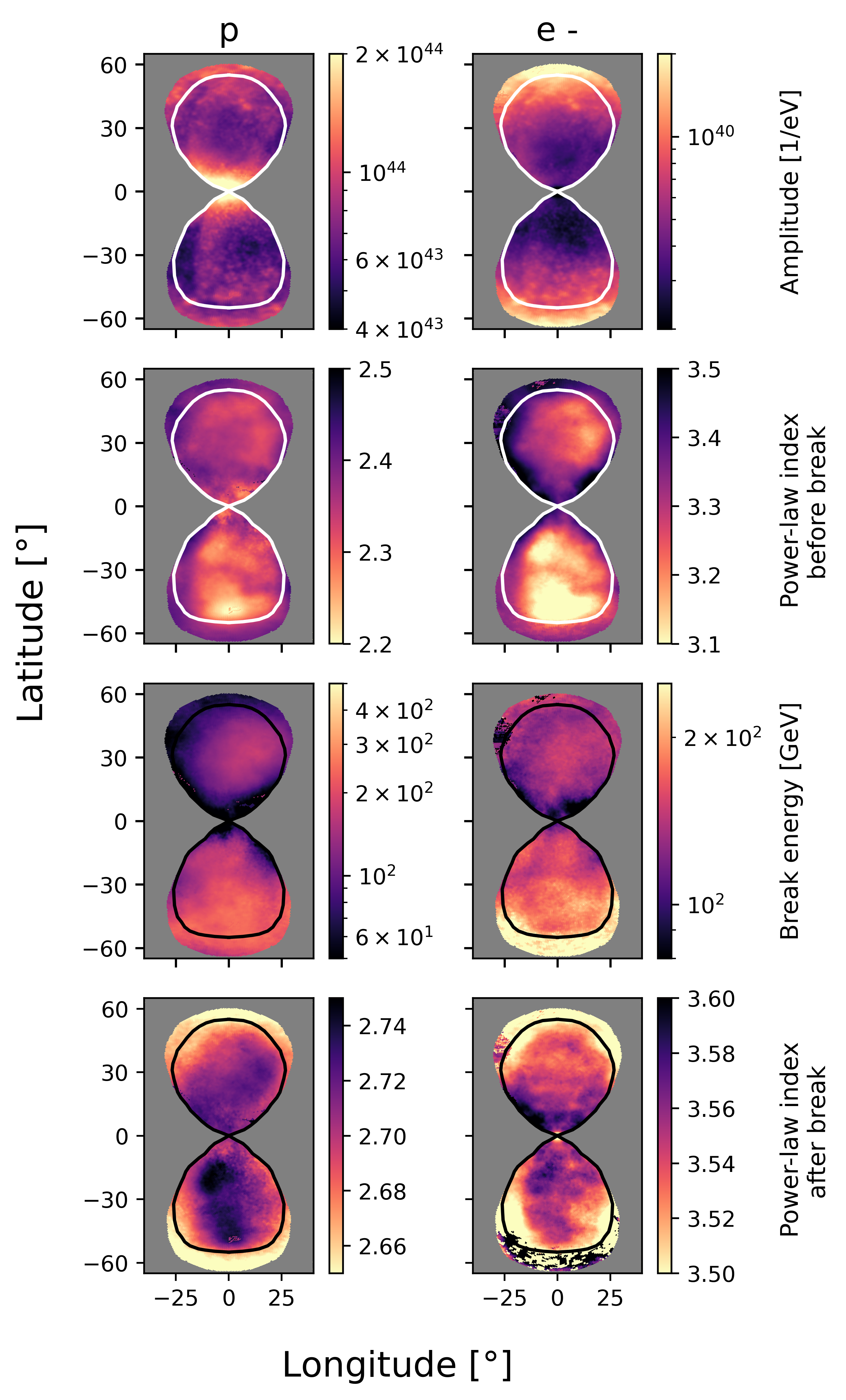}
\caption{Best-fit parameters for the BPL models (protons on the left and electrons on the right). The top panel shows the model amplitude. The second and fourth panel show the spectral index before and after the break respectively. The break energy does not show much variation along the length of the FBs as shown in the third panel.}
\label{fig:bpl_model}
\end{figure}

We start by discussing how well each model fits the data. \autoref{fig:model_comparison} shows the value of $\log{\chi^2_\mathrm{red}}$ for our six models -- three functional forms (PL, EPL, BPL) times two particle types (protons and electrons) -- in every pixel. We remind readers that we do not expect values around unity for well-fitting models because the uncertainties are correlated, and so the main quantity of interest in this feature is the relative value of $\chi^2_\mathrm{red}$, not the absolute value. With this caveat in mind, we find that in general the BPL and EPL models are noticeably better fits within the FBs than a simple PL model for both electrons and protons, but that there are no significant differences in fit quality between hadronic and leptonic models.

We show a typical fit in \autoref{fig:fit_comp}, which compares the best-fit model spectra to the data for the pixel at $b=45\degree$ and $l=0.38\degree$ (labeled as `a' in \autoref{fig:gamma_ray_data}), a point within the FBs. We see that the $\gamma$-ray data are not well accommodated by radiation from power-law distributions of either protons or electrons.
We henceforth discard PL models. 
On the other hand, our results do not obviously prefer EPL over BPL models or vice versa, nor do they show obvious preferences for hadronic versus leptonic models; we therefore discuss best-fit parameters for all these cases next.

\subsection{Inferences from acceptable models}

\autoref{fig:EPL_model} shows best-fit amplitudes, power-law indices, and cutoff energies for the EPL models. Not unexpectedly, best-fit CRp models have harder spectra than best-fit  CRe models. Best-fit spectra harden towards the tip of the Southern bubble for all models. This mirrors a  spectral hardening that was previously found in the \textit{Fermi} $\gamma$-ray data \citep{yang_fermi_2014,selig_denoised_2015,platz_multi-component_2023}. Cutoff energies are of the order of a few TeV within the FBs for both hadronic and leptonic cases. In contrast to  spectral indices, best-fit break energies do not show much variation with latitude inside the FBs. %Unlike in the projected interior, outside the FBs, break energies are not robustly determined. In particular, for regions outside the FBs best-fit cutoff energies tend to very high values and fits return a large standard deviations, consistent with the $\gamma$-ray sky outside the FBs being well explained by simple power law distributions of either electrons or protons over the 1-200 GeV energy range. 
Best-fit model amplitudes increase with latitude for CRe models while, in contrast, they are maximum near the Galactic plane for  CRp models. We will shortly see that the particle energy density follows a similar pattern as the amplitude. However, we caution that the trend for the CRp models is dependent on our assumption of a constant target proton number density, and would change if we were to instead assume a density gradient \citep[as informed by a hydrodynamic or MHD model; e.g.,][]{Tourmente2025}.
%the particles might favor a simple PL model.

\begin{figure}
\centering
\includegraphics[width=0.95\linewidth]{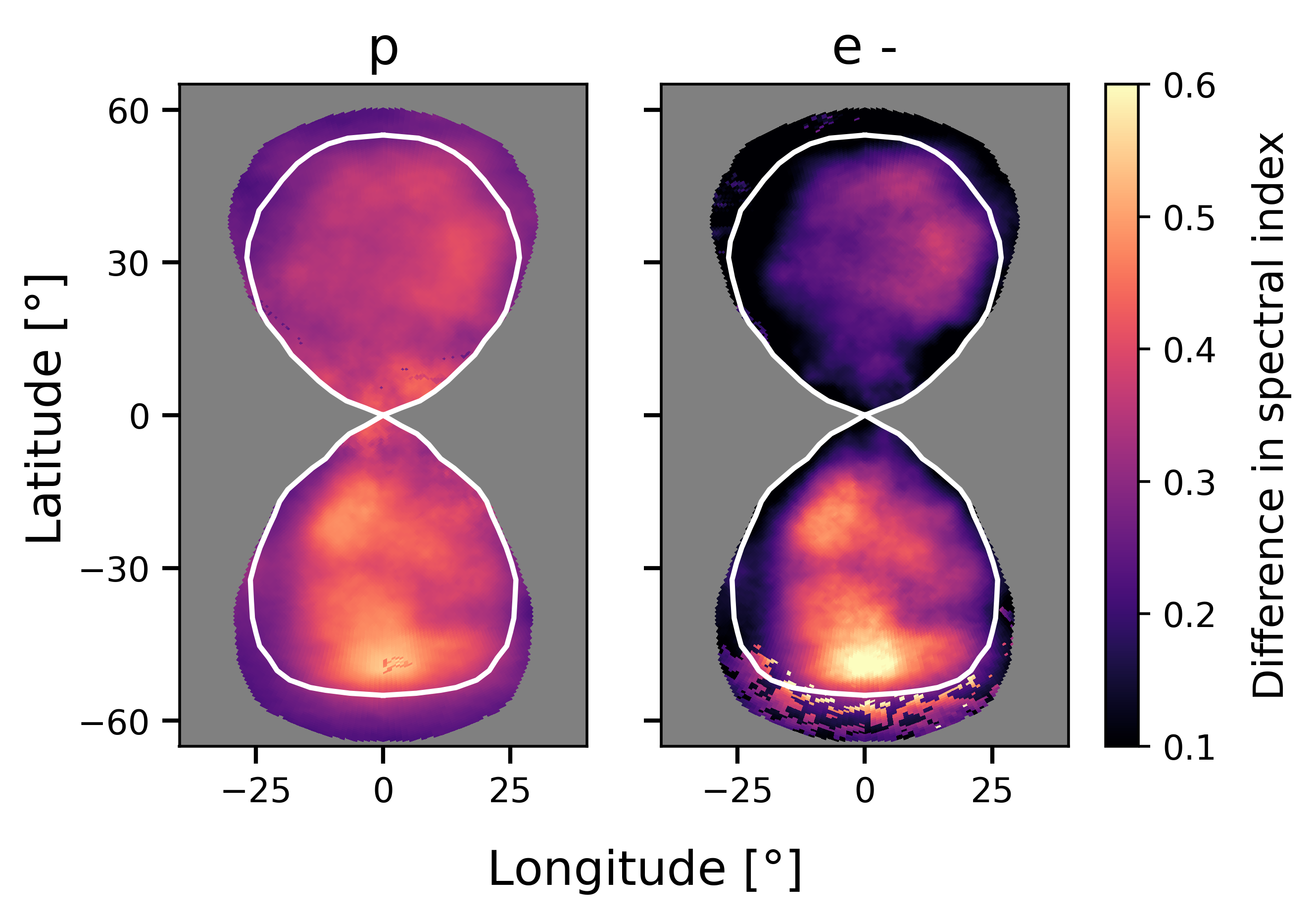}
\caption{The difference between spectral indices between the two powerlaw components in the BPL model (protons on left and electrons on right).\\}
\label{fig:diff_index}
\end{figure}

\autoref{fig:bpl_model} shows the corresponding best-fit parameters for the BPL models. Here, we find break energies of the order of $\sim$100 GeV for both cases with little latitudinal variation. Best fit spectral indices below the break are more or less consistent with spectral indices in the EPL model showing qualitatively the same hardening for large negative latitudes in the southern Bubble. In contrast, spectral indices after the break do not show variation with latitude. \autoref{fig:diff_index} shows the spatial variation of the difference between spectral indices before and after the break.
% The difference is almost zero outside the Bubble boundary for the electrons, again indicating that, while CRe models require a break inside the FBs, outside, simple PL models adequately explain the data.%
Within the bubbles, the difference in spectral index varies from $\approx 0.1 - 0.6$. Interestingly, the fits strongly rule out spectral index changes as large as unity, the value that would be expected for a synchrotron or inverse Compton cooling break -- this will become important for our discussion of the implications of our findings below.

\begin{figure*}
\centering
\includegraphics[width=0.9\textwidth]{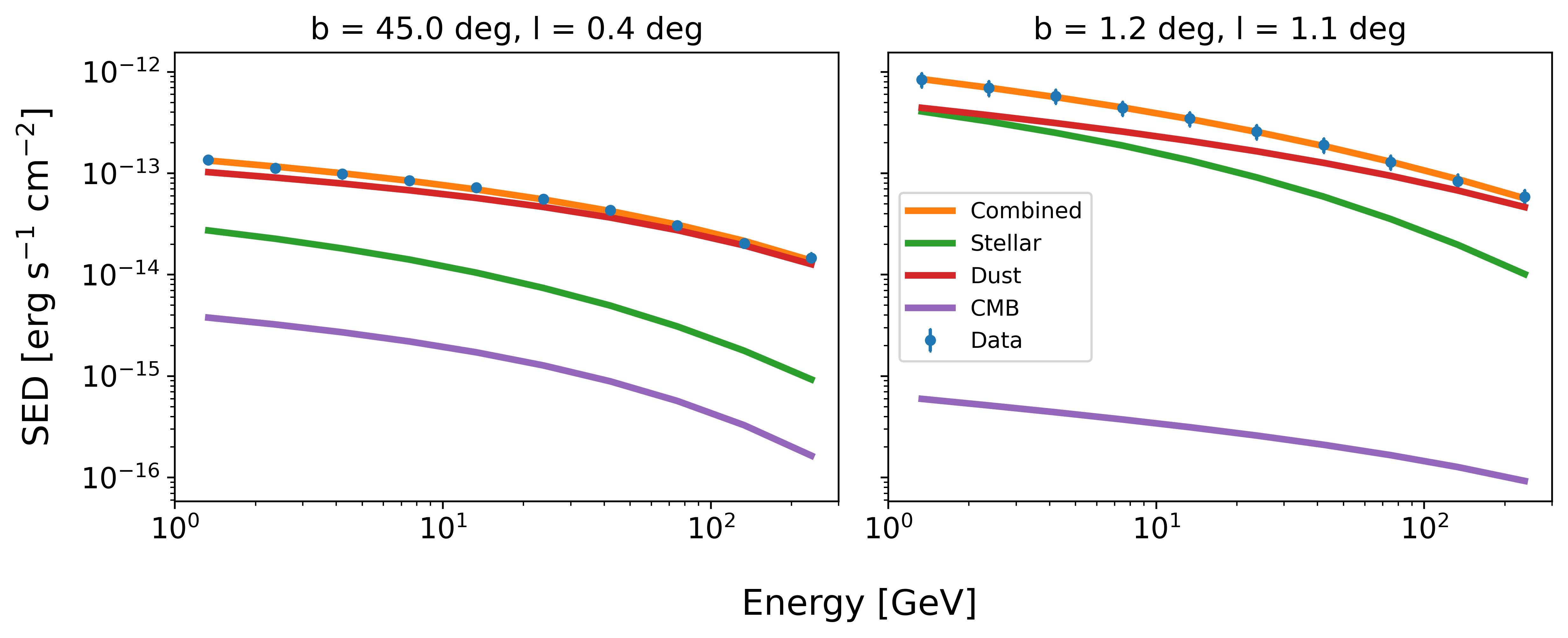}
\caption{Contributions of the stellar (green), dust (red), and CMB (purple) components of the ISRF to the total $\gamma$-ray emission (orange best-fit line) for the best-fitting leptonic model at the locations marked `a' and `b' in \autoref{fig:gamma_ray_data}. Blue points with error bars show the observed $\gamma$-ray emission. The dust component has the highest contribution at all locations, followed by the stellar component (which is suppressed by the Klein-Nishina effect); the CMB component is always sub-dominant.\\}
\label{fig:ISRF_contribution}
\end{figure*}

One important byproduct of our fits, which will be important for our interpretation, is an understanding of which radiation field components drive which parts of the observed $\gamma$-ray spectrum in leptonic models. \autoref{fig:ISRF_contribution} shows the  contribution of the different components of the ISRF -- stellar, dust, and CMB -- to the total $\gamma$-ray emission at two points labeled `a' and `b' in \autoref{fig:gamma_ray_data}. Since the stellar and dust components both contribute about equally at $\mu$m wavelengths, for this purpose we roughly divide them at 3 $\mu$m, i.e., we define the stellar component as the portion of the ISRF at wavelengths $<3$ $\mu$m, and the dust component as emission at $3-300$ $\mu$m, where 300 $\mu$m is roughly the wavelength at which the CMB begins to dominate. The energy-weighted mean photon energies of the stellar, dust and CMB components defined in this way are 1.6 eV, 0.1 eV, and $9.3\times 10^{-4}$ eV  respectively. 
We find that the dust component dominates at all $\gamma$-ray energies observable by \textit{Fermi}/LAT, followed by the starlight component and lastly the CMB. Interestingly this is true even at position `a', where the starlight field overall has a significantly larger energy density than the dust field (cf.~\autoref{fig:ISRF}). At this location, the contribution of the stellar component is comparable to dust only at lower energies while the dust 
dominates at higher energies. This is because interaction between CRe and the starlight field is strongly suppressed by Klein-Nishina effects, particularly for the higher initial photon and CRe energies that drive emission towards the top of the observed $\gamma$-ray band. 
By contrast, Klein-Nishina effects are essentially never important for the much lower-energy dust photons.

We can understand the difference in behavior by computing the Klein-Nishina parameter $\Gamma_e = 4 E_i E_e / m_e^2 c^4$, where $E_i$ and $E_e$ are the initial photon and CRe energies, respectively; Klein-Nishina suppression is important for $\Gamma_e \gtrsim 1$. For the few eV photons typical of the starlight field, $\Gamma_e$ is greater than unity for CRe energies $\gtrsim 30$ GeV, well below the break or cutoff energies of our best-fitting BPL or EPL models; by contrast, for the $\mathrm{few}\times 10^{-2}$ eV photons that characterize the dust field, $\Gamma_e \lesssim 1$ for electron energies up to $\sim$TeV, above the  break energy and comparable to the cutoff energies for our best-fitting BPL and EPL models. An important implication of this finding is that we can to good approximation assume that IC losses are in the Thomson regime throughout the FBs, since the dominant losses are due to dust-produced photons rather than starlight ones. Our interpretations for the leptonic case are based on modeling the CRe spectra for a path-averaged ISRF and thus does not account for the 3D variation of the CRe spectra.

%The top left figure shows a point far away from the Galactic plane and towards the GC (labeled by `a' in \autoref{fig:gamma_ray_data}). We see that dust is the highest contributor and CMB the lowest. The top right figure shows a point away from the Galactic plane as well as from the GC (labeled by `b' in \autoref{fig:gamma_ray_data}). Here we see that CMB begins to dominate over the stellar component for higher energies. This is because the high energy photons are in the Klein-Nishina regime and the IC scattering cross-section is suppressed. The bottom left figure shows a point near the Galactic plane as well as the GC (labeled as `c' in \autoref{fig:gamma_ray_data}). Here we see that the dust and stellar components dominate and the CMB contribution is relatively less important. The bottom right figure shows a point near the Galactic plane but away from the GC (labeled by `d' in \autoref{fig:gamma_ray_data}). Here, as well, we see that dust contribution is maximum. Hence, we conclude that emission from the dust component dominates over other fields everywhere, followed by the stellar component and then by the CMB.% 
%CMB dominates over the stellar component for higher energies.

\section{Discussion}
\label{sec:Discussion}

We next discuss some of the implications of our results, as well as comparing to earlier work.

\subsection{The energy budget of the Fermi bubbles}
\label{ssec:energy_budget}

To understand the implications of our analysis for the origins of the FBs, a useful first step is to estimate the total energies and energy densities of CRs inside the FBs in either the hadronic or leptonic scenarios. Such an estimate will necessarily be approximate: our best-fitting powerlaw indices at low energy are steeper than two, and thus the total energy depends on the minimum energy below which the powerlaw breaks. We can obtain only a weak upper limit on the presence of such a break from the fact that no break is visible in the $\gamma$-ray data, for which our lowest energy bin is $E_\gamma = 1$ GeV. In a hadronic model, $\gamma$-rays of this energy will be produced primarily by protons of $\approx 5-6$ GeV \citep{kafexhiu_parametrization_2014}. In a leptonic model where the $\gamma$-rays come primarily from IC scattering in the Thomson regime, the maximum $\gamma$-ray energy $E_{\gamma,\mathrm{max}}$ that can be produced by up-scattering of a photon with initial energy $E_i$ by a CR electron of energy $E_e$ is $E_{\gamma,\mathrm{max}} = 4 E_i (E_e/m_c c^2)^2$ \citep{Blumenthal70a}; using $E_{\gamma,\mathrm{max}} = 1$ GeV and $E_i \approx 0.03$ eV, typical of our dominant dust-produced photon field, gives $E_e \approx 45$ GeV. Given the weakness of these limits, and the fact that they are different between the leptonic and hadronic models, we instead elect to calculate the energy budgets for a nominal minimum CR energy $E_\mathrm{min} = 1$ GeV. For our measured powerlaw indices -- $\alpha \approx \alpha_1 \approx 2.2-2.3$ for protons, and $\approx 3$ for electrons -- our estimate for the total energy in a hadronic model is weakly dependent on this choice, $E_{\mathrm{tot}} \propto E_\mathrm{min}^{-k}$ with $k\approx 0.2-0.3$, while our estimate in the leptonic case is more strongly dependent, $k\approx 1$. On the other hand, in the hadronic case the total energy will scale with our assumed number density as $E_\mathrm{tot} \propto n_\mathrm{H}^{-1}$.

\begin{figure}
\centering
\includegraphics[width=0.95\linewidth]{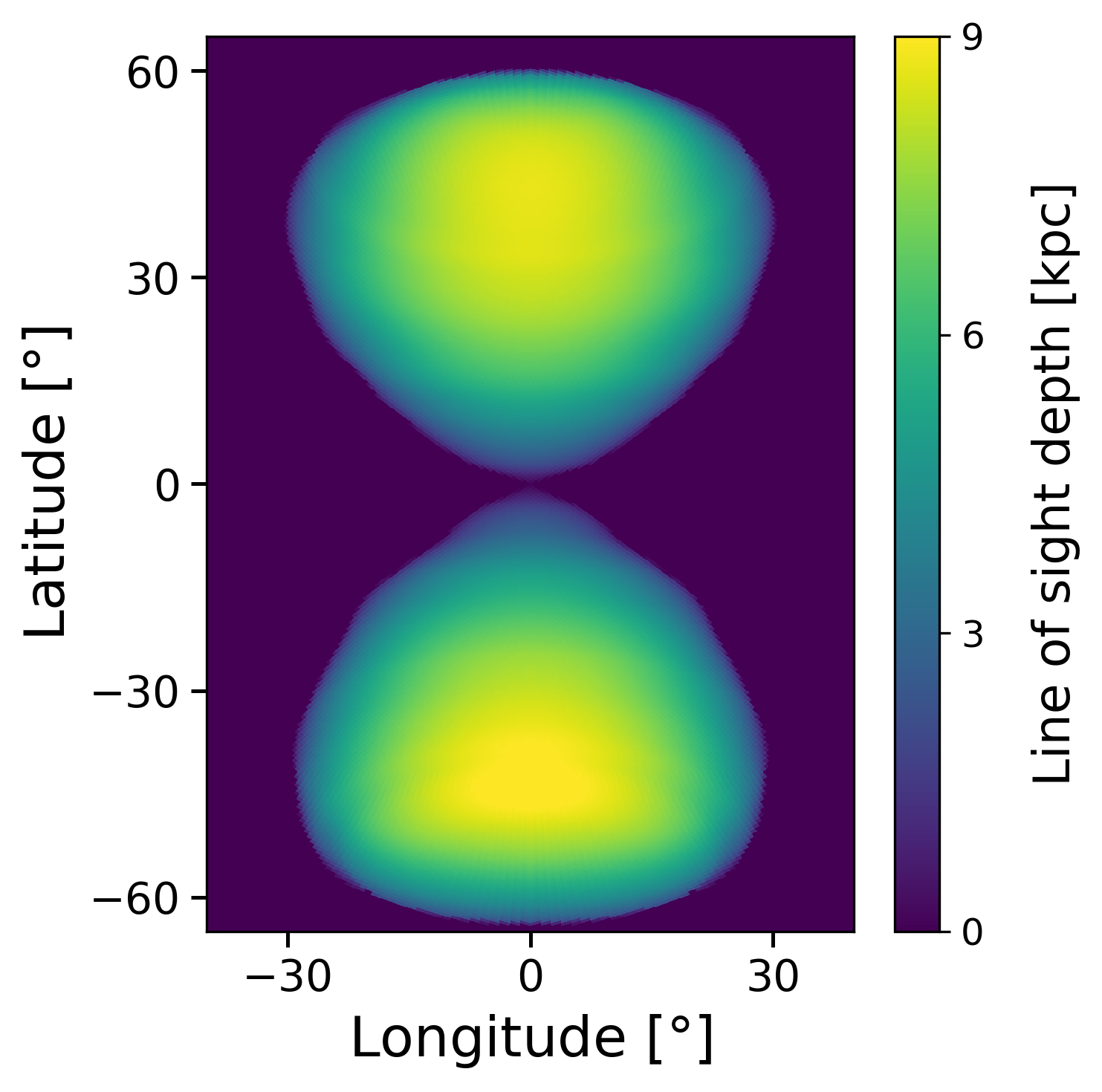}
\caption{The line of sight depth of each pixel $d_\mathrm{(\ell,b)}$, calculated by assuming the 3D structure of the FBs formed by rotating their boundary about the $z$-axis. See \autoref{ssec:isrf_density} for details.\\}
\label{fig:LOS_depth}
\end{figure}

\begin{figure*}
\centering
%\begin{subfigure}{0.45\textwidth}
\includegraphics[height=0.35\textheight]{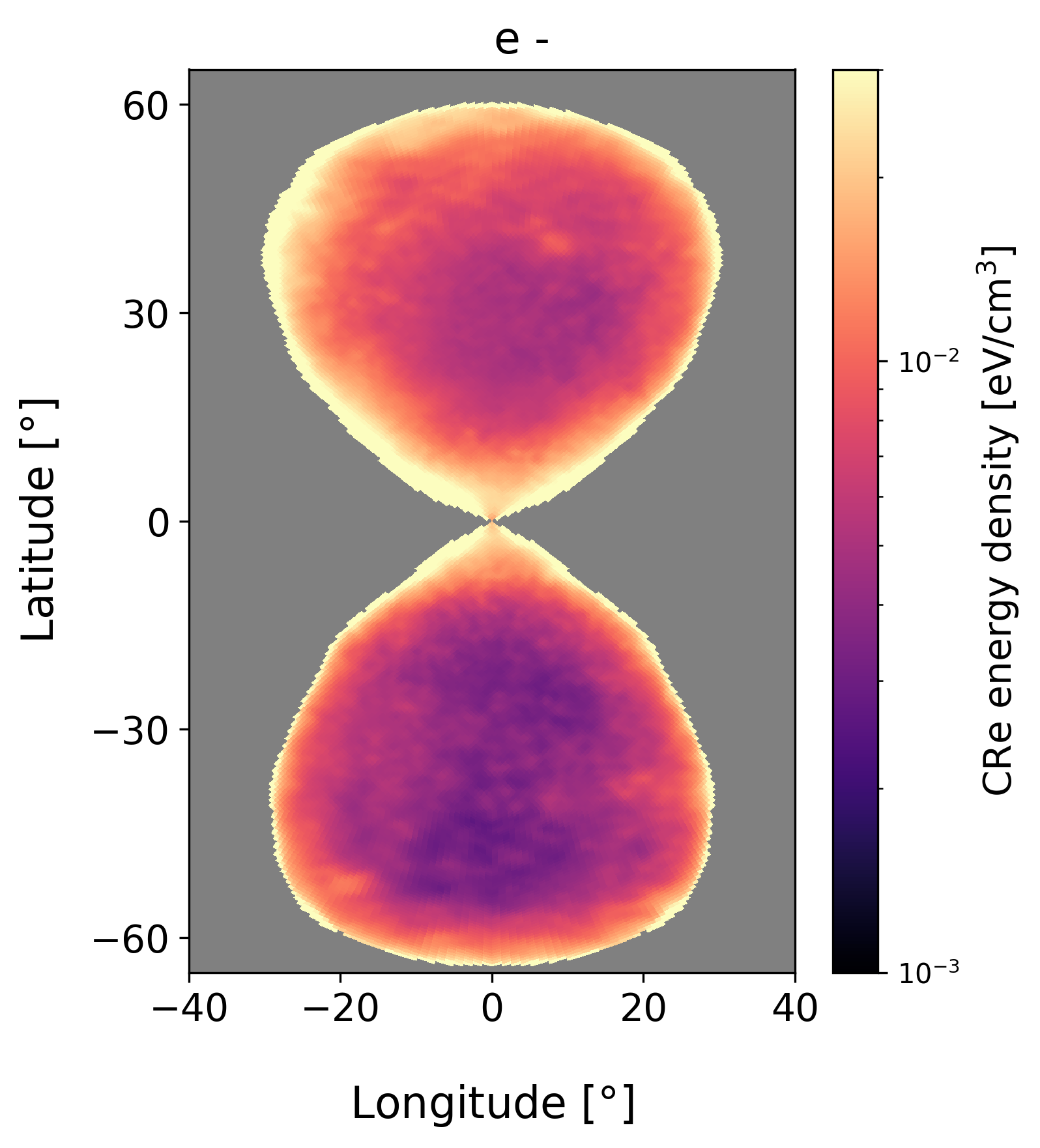}
%\end{subfigure}
%\begin{subfigure}{0.45\textwidth}
\includegraphics[height=0.35\textheight]{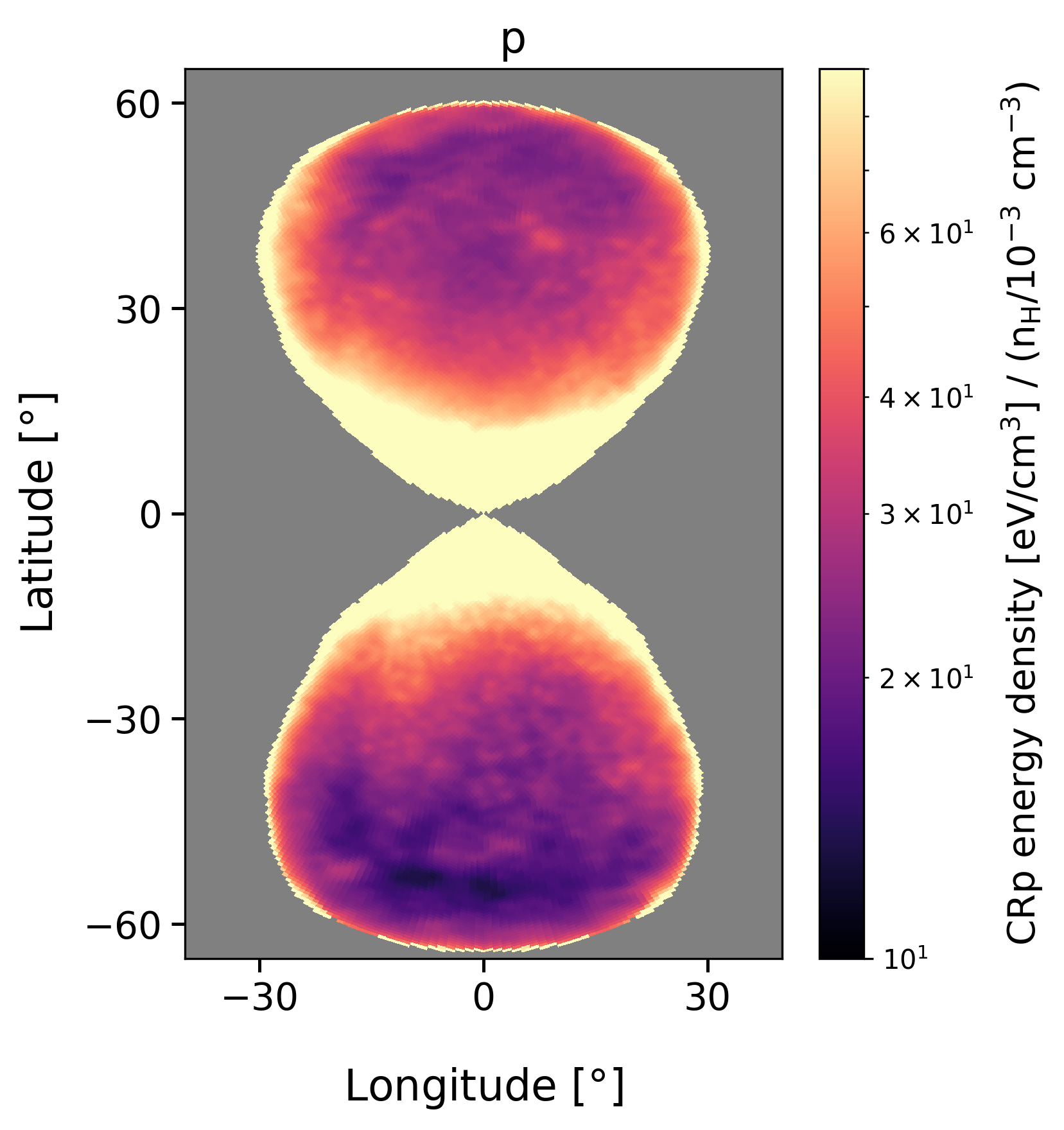}
%\end{subfigure}
\caption{CR energy density inferred for the best-fitting leptonic (left) and hadronic (right) EPL models. Note the normalization of the CRp population with respect to the fiducial value of the ambient hydrogen number density ($n_\mathrm{H}$).\\}
\label{fig:en_den}
\end{figure*}

We use these assumed lower limits together with our best-fit model parameters in the EPL and BPL models to compute the total CR energy in each pixel, and we sum over all the pixels inside the FBs to obtain total energies of $1.7 \times 10^{57}/n_\mathrm{H,-3} $ erg, $1.6 \times 10^{57}/n_\mathrm{H,-3} $ erg, 3.6 $\times 10^{53} $ erg, and 3.3 $\times 10^{53} $ erg 
%\textbf{AMI, FILL IN NUMBERS}%
in the hadronic EPL, hadronic BPL, leptonic EPL, and leptonic BPL scenarios, respectively, where $n_\mathrm{H,-{3}}$ is the number density of H nuclei normalized by 0.001 cm$^{-3}$. We also estimate the CR energy density within the FBs. To do this, we make use of the line of sight depths for each pixel $d_{(\ell,b)}$ computed in \autoref{ssec:isrf_density}; we show our calculated depths in \autoref{fig:LOS_depth}. 
%For this purpose, we assume that the FBs are cylindrically-symmetric, and compute thickness of every pixel by rotating the projected boundary shown in \autoref{fig:gamma_ray_data} about the $z$-axis.
We treat each pixel as a prism with this thickness and an area equal to its area on the sky computed at a distance equal to the distance between the Sun and the pixel (\autoref{ssec:isrf_density}), and from this plus the total energy content of the pixel as deduced from our fits we compute the energy density. We show the result in \autoref{fig:en_den} for the EPL model; the results for the BPL model are qualitatively similar. 

We can see that the inferred energy densities follow the pattern of the model amplitudes in \autoref{fig:EPL_model}.
Here we caution that CRp energy density plot should not be over-interpreted given we adopt a fixed target gas density (while the true ambient gas number density distribution is likely strongly position-dependent). That said, it is interesting to compare the pressures implied by the results in \autoref{fig:en_den} to gas pressures. For our nominal $n_\mathrm{H} = 10^{-3}$ cm$^{-3}$, the CR pressure is one third of the value shown in the right panel of this figure; the range we find, $U_\mathrm{CR} \approx (10-30) n^{-1}_{\mathrm{H},-{3}}$ eV cm$^{-3}$, corresponds to $P_\mathrm{CR}/k_\mathrm{B} \approx 4-10)n^{-1}_{\mathrm{H},-3}\times 10^4$ K cm$^{-3}$, while the pressure of fully-ionized\footnote{Meaning that there are 2.3 free particles per H nucleon for the standard He/H ratio.} thermal gas with temperature $T$ is $P_\mathrm{th}/k_\mathrm{B} = 2.3 n_\mathrm{H}T = 2.3n_{\mathrm{H},-{3}} T_6\times 10^{3}$ K cm$^{-3}$, where $T_6 = T/(10^6\,\mathrm{K})$. Thus for our nominal $n_\mathrm{H}$, and adopting $T_6 \approx 4-5$ \citep{2016ApJ...829....9M}, CR pressure is likely dominant over thermal gas pressure, though only by a factor of a few. However, since $P_\mathrm{CR}/P_\mathrm{th}\propto n_{\mathrm{H}}^{-2}$, this conclusion is strongly dependent on our choice of fiducial $n_\mathrm{H}$, and could vary in either direction for different choices.
%CR pressure will dominate if the gas density is lower than our fiducial estimate.

%
For leptonic models we can be considerably more confident about the variation of energy density with position, since we use a self-consistent, spatially-dependent ISRF model. For this case, a striking  feature is that the CRe energy density peaks towards the edges of the FBs, qualitatively consistent with limb brightening, one interpretation of which is emission from a shell structure (cf.~\citealt{su_giant_2010}). Thus not only does the CRe spectrum harden toward the FB edges, the total energy content increases as well.

\subsection{Cooling timescales}

In addition to energy budgets, we can also calculate characteristic cooling timescales for the non-thermal particle population powering the FBs, and the combination of these two provides useful constraints on their possible origin. For the hadronic case, we calculate the luminosity at each pixel by multiplying the observed intensity of the non-dust-associated emission component by the area of each pixel and numerically integrating over energy bins. For our assumed geometry of the FBs, this gives a total luminosity of $4 \times 10^{37} $ erg s$^{-1}$. The statistical error on this is very small (order $10^{33}$ erg s$^{-1}$) due to the large number of pixels, but as noted above the dominant errors are systematics in the decomposition of the $\gamma$-ray sky into components, which render the errors highly-correlated; thus a more realistic uncertainty estimate is of order of the sizes of the correlated error bars, which is tens of percent.
We have to apply two corrections to this luminosity
for full energy accounting.
First, we have to make a bolometric correction $\sim 2$ (given the inferred CRp distribution) to account for production of $\gamma$-rays above and below our sensitivity band.
Second,
given that hadronic $\gamma$-ray emission involves production of approximately equal numbers of $\pi^+$ and $\pi^-$ and $\pi^0$,
we have to factor in a further factor of $\sim$3 to correct for the power ultimately liberated in neutrinos and secondary electrons and positrons in addition to gamma-rays.
We find, therefore, 
that the total $pp$
power loss within the the FBs is $ \approx 2.4 \times 10^{38}$ erg s$^{-1}$. To estimate the time required for the current CR proton population to cool under this loss, we compute the CRp energy in protons with energies $>5.6$ GeV -- the energy range that contributes to the observable emission -- using the same method as described in \autoref{ssec:energy_budget}. Doing so gives a total energy budget of $7.9 \times10^{56}$ erg %\textbf{AMI FILL IN}%
over this energy range, and thus a characteristic hadronic cooling timescale
%\begin{equation}
%    t_{\rm had} \sim \frac{XXX\;\mathrm{erg\;s}^{-1}}{XXX n_{\mathrm{H},-2}^{-1}\,\mathrm{erg}} = 
%    XXX n_{\mathrm{H},-2}^{-1}\;\mathrm{Myr}
%\end{equation}%
\begin{equation}
    t_{\rm had} \sim \frac{7.9 \times 10^{56}n_{\mathrm{H},-3}^{-1}\,\mathrm{erg}}{2.4\times 10^{38}\;\mathrm{erg\;s}^{-1}} \approx100 \, n_{\mathrm{H},-3}^{-1}\;\mathrm{Gyr}
\end{equation}
Note that this cooling time is equivalent to the time to reach saturation at the given $n_\mathrm{H}$ defined by \citet{crocker_fermi_2011}, and that, while the absolute energies we have quoted depend on the geometry of the bubbles, the cooling timescale does not, since the numerator and denominator both depend on distance in the same way.

\begin{figure}
    \centering
    \includegraphics[width=0.9\linewidth]{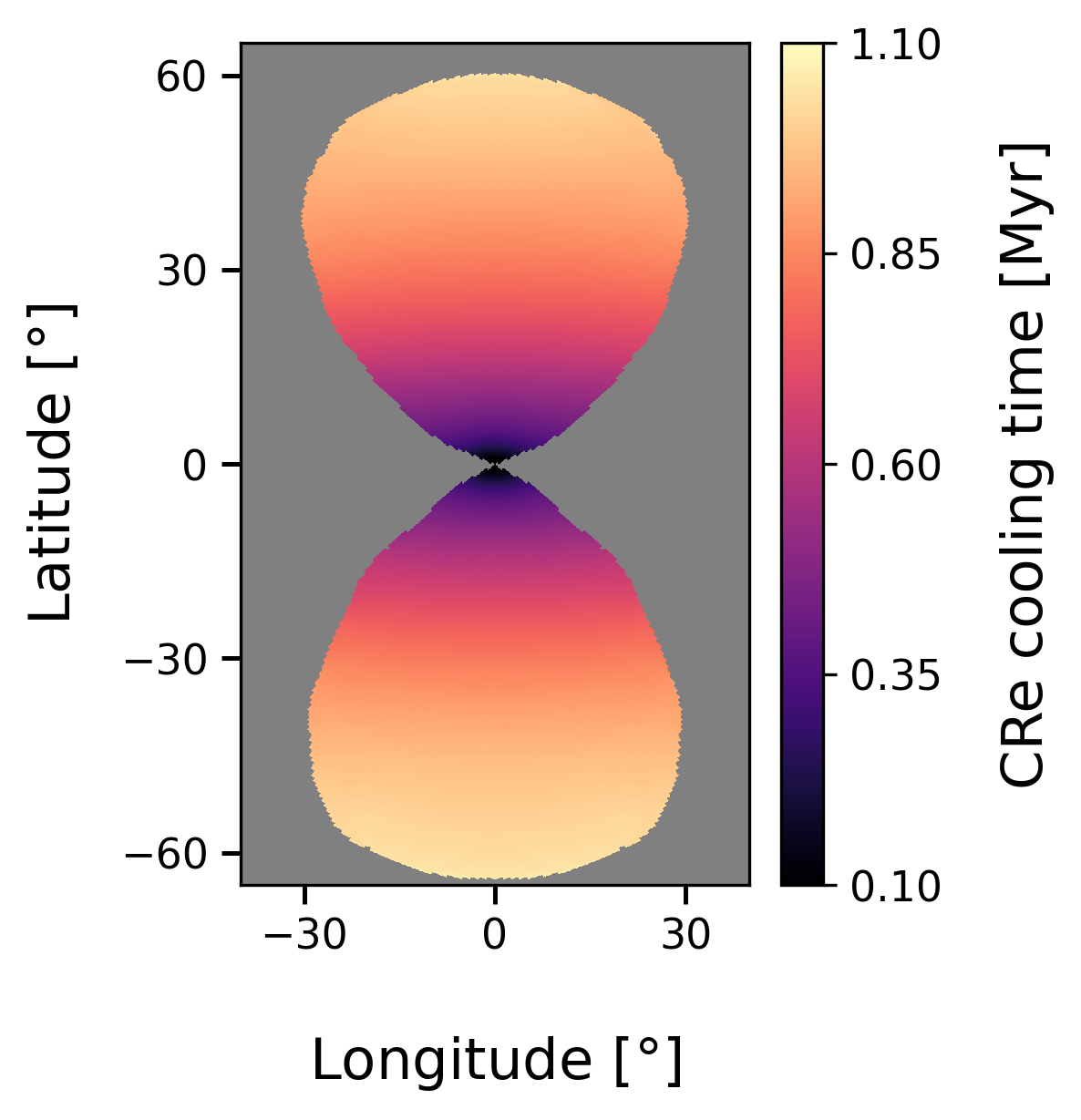}
    \caption{Inverse Compton cooling time of CRe with energy $E_\mathrm{cut}$.\\}
    \label{fig:cooling time}
\end{figure}

For leptonic models, given the strong decrease in cooling time with particle energy, it is more useful to estimate the cooling time for the highest-energy electrons than for the population as a whole. The question therefore becomes how to identify the highest energies that contribute to the observed emission, which is model-dependent. In the BPL model, we cannot identify the break energy with a cooling break, since the difference in spectral indices between the upper and lower powerlaws is $\lesssim 0.5$, far smaller than the value of unity expected for an IC cooling break in the Thomson regime (cf.~\autoref{fig:diff_index}), and so the population must be un-cooled up to the highest energies that contribute to any of the \textit{Fermi}/LAT data. For the EPL model, we can identify the highest electron energy with the cutoff energy $E_\mathrm{cut} \approx 1-2$ TeV. Since the latter is more permissive regarding the cooling timescale, we calculate the cooling time in the EPL model. We use \texttt{GAMERA}\footnote{\url{https://libgamera.github.io/GAMERA/docs/documentation.html}} to compute the total IC energy loss rate $\dot{E}_\mathrm{IC}$ for a CRe with energy $E_\mathrm{cut}$ in each pixel, and define the cooling time as $E_\mathrm{cut}/\dot{E}_\mathrm{IC}$. We plot this cooling time in \autoref{fig:cooling time}. We emphasise that this IC cooling time actually defines an upper limit to the total loss time  given the existence of magnetic fields suffusing the Bubbles (on which CRe can synchrotron radiate) with characteristic energy densities similar to those of the dust field \citep{carretti_giant_2013}.

\subsection{Implications of the cooling timescales}

\subsubsection{Hadronic scenarios}

As has long been known \citep{crocker_fermi_2011,sarkar_fermierosita_2024}, the CRp power required to sustain the $\gamma$-ray emission from the FBs in a hadronic scenario -- $2.4 \times 10^{38}$ erg s$^{-1}$  -- is rather modest and could be sustained by the current rate of nuclear star formation, even if we have to make upwards corrections of order of a few to account for additional Coulombic/ionisation losses.
On the other hand, as has also long been appreciated, 
the timescale to reach a hadronic steady state (as set by the $pp$ loss time) is long given the low density of the volume-filling phase of the Bubbles; given that, one must either countenance that the Bubbles are very old (multi-Gyr) structures and/or quite far from steady state.
In the latter case, energy/power demands become increasingly difficult to be accommodated by star formation over a timescale smaller than the Bubbles' age or even, at the extreme, the age of the Galaxy.
For instance, above we determine the total CRp energy content of the Bubbles to be $\sim 2 \times 10^{57}$ erg$/n_{\rm H,-3}$. To put this number in perspective, if we assume that each core collapse supernova accelerates $\approx 10^{50}$ erg of CRp's and that there is one such supernova per $\approx 100$ M$_\odot$ of stars formed \citep[e.g.,][and references therein]{roth_diffuse_2021}, then our estimate of the total energy in the FBs in a hadronic scenario corresponds to the total CRp energy released by the formation of a stellar population with a mass $\sim 2 \times 10^9 \ M_\odot/n_{\rm H,-3}$.
For $n_{\rm H,-3} \approx 1$, this mass is comparable to the total mass of the nuclear bulge \citep[][and references therein]{Bland-Hawthorn16b}, a sub-structure of the Galactic bulge that, 
while a plausible morphological match to the base of the Bubbles, 
has formed over a timescale comparable to the age of the Milky Way. This would imply that the FBs are similarly old.
Of course, this constraint is relaxed to the extent that the effective target density sampled by the putative CRp's is, in reality, larger than our chosen reference value of $10^{-3}$ cm$^{-3}$, as might be the case if the CRp's are concentrated into regions of systematically higher density than the volume-filling phase \citep[cf.][]{crocker_steady-state_2014,crocker_unified_2015}.

In any case, these considerations reveal that determinations of the Bubbles' age independent of assumptions around the $\gamma$-ray production microphysics are quite discriminatory for hadronic scenarios as, indeed, would be reliable determinations of the effective $n_{\rm H}$ sampled by CRp's in the Bubbles.
Energetics constraints on hadronic models would be rendered even more severe in the case that CRp's suffer additional adiabatic losses in the case 
that they are rapidly advected out of the nucleus co-mingled with expanding thermal plasma and magnetic fields\footnote{Such adiabatic losses do not occur in the limit that the transport of the CRp's  is diffusive with respect to the background plasma. Even if adiabatic losses do occur, they can be largely recovered because of (re)acceleration at shocks that occur in the flow \citep[e.g.,][]{Lacki2014,crocker_unified_2015}.}.

Overall, in the case that the timescale for the formation of the Bubbles becomes much less than, say, $\sim$100 Myr, or their total CRp energy content were found to be $\gtrsim 10^{57}$ erg, nuclear or inner Milky Way star formation is increasingly strongly excluded and one is pushed towards invoking the Galaxy's supermassive black hole as their ultimate generator.
An AGN scenario origin for the Bubbles could be either leptonic or hadronic (or both), of course.

As has long been known \citep{su_giant_2010, ackermann_spectrum_2014, crocker_unified_2015} a potential point in favour of a leptonic model is that CR electrons from the same, quasi power law distribution can account for both the hard-spectrum, microwave haze emission and the gamma-ray emission. On the other hand, a single power law population does not explain \citep{crocker_unified_2015} the relatively steep spectrum radio continuum emission from the S-PASS Lobes \citep{carretti_giant_2013}.

\subsubsection{Leptonic scenarios}

For the case that the $\gamma$-ray emission from the Bubbles is primarily leptonic, we find that one needs an increasingly large CRe energy density towards the Bubbles' edges; {\it a priori} this seems hard to accommodate within scenarios where one advects CRe's out of the nucleus, but it can be accommodated accounting for non-steady state effects as demonstrated, e.g., by the model of \citet{yang_spatially_2017}.
Probably the most severe test of any scenario invoking leptonic emission from advected, nuclear CRe's is the short cooling time of electrons radiating at $\sim$200 GeV.
This is $\lesssim$ 1 Myr at the top of the Bubbles accounting only for IC losses; factoring in synchrotron losses (certainly present at some level), bremsstrahlung and ionisation (important at lower latitudes/earlier in the expansion for an explosion), even stronger IC losses in the higher energy density fields closer to the plane (through which the CRe's must pass), and adiabatic losses (which, again, should be important in an explosion), the real, effective loss time is likely  significantly smaller.

Transport to 8 kpc (the top of the Bubbles) within 1 Myr implies an expansion speed $\sim$8000 km/s, or $\sim 0.03c$, a highly supersonic flow that seems difficult to reconcile with the much lower Mach numbers of expansion inferred from X-ray phenomenology near the Bubbles' edges \citep[see the extended discusion in][]{sarkar_fermierosita_2024}.
Overall, to accommodate these transport time constraints and the peaking of the putative CRe's energy density towards or at the Bubbles' edges, it seems more natural to us -- as it has also seemed to many previous authors \citep[][and references therein]{sarkar_fermierosita_2024} -- that there be distributed, {\it in situ} acceleration concentrated towards the Bubbles' periphery, however that is achieved.  

\section{Summary and conclusions} 

In this work, we investigate the cosmic ray distributions required to produce the measured $\gamma$-ray emission in the Fermi Bubbles (FBs) on a pixel-by-pixel basis, leveraging a recent template-free reconstruction of \textit{Fermi} data by \cite{platz_multi-component_2023} togther with a modern model for the position-dependent interstellar radiation field from \citet{popescu_radiation_2017}. We consider both leptonic models in which IC scattering is the primary emission mechanism and hadronic ones in which pion production dominates, and derive pixel-by-pixel fits for the cosmic ray electron (CRe) or proton (CRp) spectra over 20,000 pixels, thereby tracing out the variations in these spectra across the FBs.

We show that simple power law models for both CRe and CRp energy distributions predict $\gamma$-ray spectra that provide poor fits to the data, enabling us to rule them out, but that power laws with exponential cutoffs and broken power law models both provide similarly good fits to the data; the fits are indistinguishably good for both CRe and CRp scenarios. Best-fit spectral indices at the low-energy end for either class of model show hardening towards the caps of the FBs, particularly in the South, for both CRp and CRe models. %This result \textcolor{red}{is in contrast} with the earlier findings of \citet{narayanan_latitude-dependent_2017}, who bin the data by latitude  and do not find any spectral hardening \textcolor{red}{in the CRe spectrum. We note that their analysis extends only to b = -35$\degree$ latitude, while the spectral hardening reported here becomes prominent at higher Southern latitudes beyond this range}. 
Note that while this result seems to be in tension with the earlier findings of \citet{narayanan_latitude-dependent_2017}, who bin the data by latitude  and do not find any spectral hardening in the CRe spectrum,
their analysis extends only to $b = -35\degree$ latitude,
below the latitudes where we find the spectral hardening reported here. 
Best-fit cutoff energies in exponential cutoff models are of the order of a few TeV, while break energies in broken powerlaw models are a few hundred GeV; neither exhibit much variation with latitude. The change in spectral index for broken powerlaw models is $\lesssim 0.5$, inconsistent with an inverse Compton or synchrotron cooling break.

%We compare our results with the results obtained by \cite{narayanan_latitude-dependent_2017}, who have also carried out a latitude dependent analysis of the FBs by fitting a CRe distribution model to the gamma-ray and microwave data. While our results are consistent with their conclusion that the CRe have high cut-off energies, we find a difference in the spectral index behaviour. We find that the spectral index hardens towards the Southern tip of the Bubble, while their results show no variation of the CRe spectrum inferred from the gamma ray data. 

Our fits allow us to calculate the total energy, the position-dependent energy densities, and the cooling times for CRs in the FBs. While these are significantly uncertain for hadronic models due to the unknown distribution of thermal background gas, for leptonic models -- where the uncertainty is smaller because the radiation field is better known -- the fits impose important constraints. The data require the CRe energy density to increase with distance from the plane, particularly towards the edge of the bubbles, with a variation as a function of position that is morphologically similar to the spectral hardening. Moreover, the cooling times are extremely short -- $\approx 1$ Myr at the bubble caps, but falling to $\approx 0.1$ Myr near the Galactic center. These findings place strong constraints on the models where the CRe driving the emission are injected at the nucleus and then transported to the bubble edge, implying either that CRe must be injected at relativistic speeds (as in the jet driven scenario \citealt{yang_spatially_2017}), or\textit{in situ} acceleration of leptonic CRs at the bubble caps. Alternatively, the observations could be explained by a hadronic scenario that does not suffer from such severe cooling limits.
\\

\section*{Acknowledgments}

AT acknowledges the support of the ANU Future Research Talent (FRT) program, which provided the funding and the opportunity to undertake part of this research at the Australian National University. RMC and MRK acknowledge support from the Australian Research Council through the \textit{Discovery Projects} scheme, award DP230101055. This research was undertaken with the assistance of resources from the National Computational Infrastructure (NCI Australia), an NCRIS enabled capability supported by the Australian Government, through award jh2.

\section*{Data Availability}

Our best-fitting parameters for the leptonic and hadronic EPL and BPL models for all pixels are available from \url{https://doi.org/10.5281/zenodo.19675422}.

%\bibliographystyle{apsrev4-1}
% You should give the same name for your .bbl as your main .tex
% since it is a requirement for posting on ArXiv.
\bibliographystyle{aasjournal}
\bibliography{Fermi_bubble}

\end{document}